\documentclass[twocolumn,letterpaper,aps,prd,superscriptaddress,nofootinbib,showpacs,floatfix]{revtex4}

\usepackage{graphicx}	
\usepackage{xspace}     

\begin{document}


\title{Direct-Photon Production in $p$$+$$p$ Collisions at $\sqrt{s}=200$ GeV 
at Midrapidity}

\newcommand{\abilene}{Abilene Christian University, Abilene, Texas 79699, USA}
\newcommand{\acadsin}{Institute of Physics, Academia Sinica, Taipei 11529, Taiwan}
\newcommand{\banaras}{Department of Physics, Banaras Hindu University, Varanasi 221005, India}
\newcommand{\barc}{Bhabha Atomic Research Centre, Bombay 400 085, India}
\newcommand{\bnlcoll}{Collider-Accelerator Department, Brookhaven National Laboratory, Upton, New York 11973-5000, USA}
\newcommand{\bnlphys}{Physics Department, Brookhaven National Laboratory, Upton, New York 11973-5000, USA}
\newcommand{\caucr}{University of California - Riverside, Riverside, California 92521, USA}
\newcommand{\charlesczech}{Charles University, Ovocn\'{y} trh 5, Praha 1, 116 36, Prague, Czech Republic}
\newcommand{\ciae}{Science and Technology on Nuclear Data Laboratory, China Institute of Atomic Energy, Beijing 102413, P.~R.~China}
\newcommand{\cns}{Center for Nuclear Study, Graduate School of Science, University of Tokyo, 7-3-1 Hongo, Bunkyo, Tokyo 113-0033, Japan}
\newcommand{\colorado}{University of Colorado, Boulder, Colorado 80309, USA}
\newcommand{\columbia}{Columbia University, New York, New York 10027 and Nevis Laboratories, Irvington, New York 10533, USA}
\newcommand{\czechtech}{Czech Technical University, Zikova 4, 166 36 Prague 6, Czech Republic}
\newcommand{\dapnia}{Dapnia, CEA Saclay, F-91191, Gif-sur-Yvette, France}
\newcommand{\debrecen}{Debrecen University, H-4010 Debrecen, Egyetem t{\'e}r 1, Hungary}
\newcommand{\elte}{ELTE, E{\"o}tv{\"o}s Lor{\'a}nd University, H - 1117 Budapest, P{\'a}zm{\'a}ny P. s. 1/A, Hungary}
\newcommand{\fit}{Florida Institute of Technology, Melbourne, Florida 32901, USA}
\newcommand{\fsu}{Florida State University, Tallahassee, Florida 32306, USA}
\newcommand{\gsu}{Georgia State University, Atlanta, Georgia 30303, USA}
\newcommand{\hiroshima}{Hiroshima University, Kagamiyama, Higashi-Hiroshima 739-8526, Japan}
\newcommand{\ihepprot}{IHEP Protvino, State Research Center of Russian Federation, Institute for High Energy Physics, Protvino, 142281, Russia}
\newcommand{\illuiuc}{University of Illinois at Urbana-Champaign, Urbana, Illinois 61801, USA}
\newcommand{\inrras}{Institute for Nuclear Research of the Russian Academy of Sciences, prospekt 60-letiya Oktyabrya 7a, Moscow 117312, Russia}
\newcommand{\instpasczech}{Institute of Physics, Academy of Sciences of the Czech Republic, Na Slovance 2, 182 21 Prague 8, Czech Republic}
\newcommand{\isu}{Iowa State University, Ames, Iowa 50011, USA}
\newcommand{\jinrdubna}{Joint Institute for Nuclear Research, 141980 Dubna, Moscow Region, Russia}
\newcommand{\kek}{KEK, High Energy Accelerator Research Organization, Tsukuba, Ibaraki 305-0801, Japan}
\newcommand{\korea}{Korea University, Seoul, 136-701, Korea}
\newcommand{\kurchatov}{Russian Research Center ``Kurchatov Institute", Moscow, 123098 Russia}
\newcommand{\kyoto}{Kyoto University, Kyoto 606-8502, Japan}
\newcommand{\labllr}{Laboratoire Leprince-Ringuet, Ecole Polytechnique, CNRS-IN2P3, Route de Saclay, F-91128, Palaiseau, France}
\newcommand{\lawllnl}{Lawrence Livermore National Laboratory, Livermore, California 94550, USA}
\newcommand{\losalamos}{Los Alamos National Laboratory, Los Alamos, New Mexico 87545, USA}
\newcommand{\lpc}{LPC, Universit{\'e} Blaise Pascal, CNRS-IN2P3, Clermont-Fd, 63177 Aubiere Cedex, France}
\newcommand{\lund}{Department of Physics, Lund University, Box 118, SE-221 00 Lund, Sweden}
\newcommand{\mass}{Department of Physics, University of Massachusetts, Amherst, Massachusetts 01003-9337, USA }
\newcommand{\muenster}{Institut f\"ur Kernphysik, University of Muenster, D-48149 Muenster, Germany}
\newcommand{\muhlenberg}{Muhlenberg College, Allentown, Pennsylvania 18104-5586, USA}
\newcommand{\myongji}{Myongji University, Yongin, Kyonggido 449-728, Korea}
\newcommand{\nagasaki}{Nagasaki Institute of Applied Science, Nagasaki-shi, Nagasaki 851-0193, Japan}
\newcommand{\newmex}{University of New Mexico, Albuquerque, New Mexico 87131, USA }
\newcommand{\nmsu}{New Mexico State University, Las Cruces, New Mexico 88003, USA}
\newcommand{\ornl}{Oak Ridge National Laboratory, Oak Ridge, Tennessee 37831, USA}
\newcommand{\orsay}{IPN-Orsay, Universite Paris Sud, CNRS-IN2P3, BP1, F-91406, Orsay, France}
\newcommand{\peking}{Peking University, Beijing 100871, P.~R.~China}
\newcommand{\pnpi}{PNPI, Petersburg Nuclear Physics Institute, Gatchina, Leningrad region, 188300, Russia}
\newcommand{\riken}{RIKEN Nishina Center for Accelerator-Based Science, Wako, Saitama 351-0198, Japan}
\newcommand{\rikjrbrc}{RIKEN BNL Research Center, Brookhaven National Laboratory, Upton, New York 11973-5000, USA}
\newcommand{\rikkyo}{Physics Department, Rikkyo University, 3-34-1 Nishi-Ikebukuro, Toshima, Tokyo 171-8501, Japan}
\newcommand{\saispbstu}{Saint Petersburg State Polytechnic University, St. Petersburg, 195251 Russia}
\newcommand{\saopaulo}{Universidade de S{\~a}o Paulo, Instituto de F\'{\i}sica, Caixa Postal 66318, S{\~a}o Paulo CEP05315-970, Brazil}
\newcommand{\seoulnat}{Seoul National University, Seoul, Korea}
\newcommand{\stonybrkc}{Chemistry Department, Stony Brook University, SUNY, Stony Brook, New York 11794-3400, USA}
\newcommand{\stonycrkp}{Department of Physics and Astronomy, Stony Brook University, SUNY, Stony Brook, New York 11794-3400, USA}
\newcommand{\subatech}{SUBATECH (Ecole des Mines de Nantes, CNRS-IN2P3, Universit{\'e} de Nantes) BP 20722 - 44307, Nantes, France}
\newcommand{\tenn}{University of Tennessee, Knoxville, Tennessee 37996, USA}
\newcommand{\titech}{Department of Physics, Tokyo Institute of Technology, Oh-okayama, Meguro, Tokyo 152-8551, Japan}
\newcommand{\tsukuba}{Institute of Physics, University of Tsukuba, Tsukuba, Ibaraki 305, Japan}
\newcommand{\vandy}{Vanderbilt University, Nashville, Tennessee 37235, USA}
\newcommand{\waseda}{Waseda University, Advanced Research Institute for Science and Engineering, 17 Kikui-cho, Shinjuku-ku, Tokyo 162-0044, Japan}
\newcommand{\weizmann}{Weizmann Institute, Rehovot 76100, Israel}
\newcommand{\wigner}{Institute for Particle and Nuclear Physics, Wigner Research Centre for Physics, Hungarian Academy of Sciences (Wigner RCP, RMKI) H-1525 Budapest 114, POBox 49, Budapest, Hungary}
\newcommand{\yonsei}{Yonsei University, IPAP, Seoul 120-749, Korea}
\affiliation{\abilene}
\affiliation{\acadsin}
\affiliation{\banaras}
\affiliation{\barc}
\affiliation{\bnlcoll}
\affiliation{\bnlphys}
\affiliation{\caucr}
\affiliation{\charlesczech}
\affiliation{\ciae}
\affiliation{\cns}
\affiliation{\colorado}
\affiliation{\columbia}
\affiliation{\czechtech}
\affiliation{\dapnia}
\affiliation{\debrecen}
\affiliation{\elte}
\affiliation{\fit}
\affiliation{\fsu}
\affiliation{\gsu}
\affiliation{\hiroshima}
\affiliation{\ihepprot}
\affiliation{\illuiuc}
\affiliation{\inrras}
\affiliation{\instpasczech}
\affiliation{\isu}
\affiliation{\jinrdubna}
\affiliation{\kek}
\affiliation{\korea}
\affiliation{\kurchatov}
\affiliation{\kyoto}
\affiliation{\labllr}
\affiliation{\lawllnl}
\affiliation{\losalamos}
\affiliation{\lpc}
\affiliation{\lund}
\affiliation{\mass}
\affiliation{\muenster}
\affiliation{\muhlenberg}
\affiliation{\myongji}
\affiliation{\nagasaki}
\affiliation{\newmex}
\affiliation{\nmsu}
\affiliation{\ornl}
\affiliation{\orsay}
\affiliation{\peking}
\affiliation{\pnpi}
\affiliation{\riken}
\affiliation{\rikjrbrc}
\affiliation{\rikkyo}
\affiliation{\saispbstu}
\affiliation{\saopaulo}
\affiliation{\seoulnat}
\affiliation{\stonybrkc}
\affiliation{\stonycrkp}
\affiliation{\subatech}
\affiliation{\tenn}
\affiliation{\titech}
\affiliation{\tsukuba}
\affiliation{\vandy}
\affiliation{\waseda}
\affiliation{\weizmann}
\affiliation{\wigner}
\affiliation{\yonsei}
\author{A.~Adare} \affiliation{\colorado}
\author{S.~Afanasiev} \affiliation{\jinrdubna}
\author{C.~Aidala} \affiliation{\mass}
\author{N.N.~Ajitanand} \affiliation{\stonybrkc}
\author{Y.~Akiba} \affiliation{\riken} \affiliation{\rikjrbrc}
\author{H.~Al-Bataineh} \affiliation{\nmsu}
\author{J.~Alexander} \affiliation{\stonybrkc}
\author{K.~Aoki} \affiliation{\kyoto} \affiliation{\riken}
\author{L.~Aphecetche} \affiliation{\subatech}
\author{J.~Asai} \affiliation{\riken}
% ...
\collaboration{PHENIX Collaboration} \noaffiliation

\date{\today}

\begin{abstract}
The differential cross section for the production of direct photons in
$p$$+$$p$ collisions at $\sqrt{s}=200$ GeV at midrapidity was measured in
the PHENIX detector at the Relativistic Heavy Ion Collider.  
Inclusive-direct photons were measured in the transverse 
momentum range from 5.5--25 GeV/$c$, extending the range beyond previous 
measurements.  Event structure was studied with an isolation criterion.
Next-to-leading-order perturbative-quantum-chromodynamics 
calculations give a good description of the spectrum. 
When the cross section is expressed versus $x_{T}$,
the PHENIX data are seen to be in agreement with measurements 
from other experiments at different center-of-mass energies.
\end{abstract}

	
\maketitle
\section{Introduction}

Direct photons are defined as photons that do not originate from hadronic 
decays.  In hadron-hadron collisions, direct photons at large transverse 
momentum ($p_T$) are predominantly produced by the fundamental 
quantum-chromodynamics (QCD) 2-to-2 hard-scattering subprocesses, 
$g+q\rightarrow \gamma+q$ and $\bar{q}+q\rightarrow \gamma +g$, where the 
former subprocess, which dominates in $p$$+$$p$ and A$+$A collisions, is 
called ``the inverse QCD Compton effect''~\cite{Fritzsch:1977eq}. This 
subprocess is one of the most important of the QCD 2-to-2 subprocesses for 
three reasons: 
\begin{enumerate}
\item the photon is the only outgoing particle in fundamental QCD 
2-to-2 subprocesses that is a single particle, which can be measured to high 
precision;
\item the scattered quark has equal and opposite transverse momentum 
to the direct-photon so that the transverse momentum of the jet from the 
fragmented quark is also precisely known (modulo $k_T$ or multisoft gluon 
effects~\cite{Adare:2010yw}); and 
\item it is directly sensitive to the gluon 
distribution function of the proton times the distribution function of 
quarks, which is precisely measured in deeply inelastic lepton-proton 
scattering.
\end{enumerate}

If both the direct photon with $p_T$ and rapidity $y_{\gamma}$ and the
away side jet at $y_J$ are detected then,  
to the extent that the $\bar{q}+q\rightarrow \gamma +g$ subprocess can
be neglected in $p$$+$$p$ collisions due to the predominance of gluons over
anti-quarks, the jet opposite to the direct photon is a 
quark~\cite{Adare:2009vd}.
Tagging jets with direct photons
provides an excellent method of studying any medium effect on the
energy or fragmentation of the outgoing quark. Furthermore, the cross
section for $g+q\rightarrow \gamma +q$ in LO
pQCD~\cite{Fritzsch:1977eq} in scattering of hadron A from hadron B
takes on the simple form   
for the reaction $A+B\rightarrow \gamma +q +X$: 
\begin{eqnarray}
\label{eq:QCDComptonfull}
\frac{d^3\sigma}{dp_T^2\, d{y_\gamma}\, d{y_J}}
& = & x_1 g_A(x_1)\, F_{\rm 2B}(x_2) \\
& & \nonumber \times\frac{\pi\alpha\alpha_s(Q^2)}{3\hat{s}^2} 
 \left(\frac{1+\cos\theta^*}{2}+\frac{2}{1+\cos\theta^*}\right ) \\
& & \nonumber  +  x_2 g_B(x_2)\, F_{\rm 2A}(x_1) \\
& & \nonumber  \times\frac{\pi\alpha\alpha_s(Q^2)}{3\hat{s}^2}
\nonumber \left(\frac{1-\cos\theta^*}{2}+\frac{2}{1-\cos\theta^*}\right),
\nonumber 
\end{eqnarray}
where the parton kinematics are fully determined by
\begin{equation}
x_1=x_T \, \frac{e^{y_\gamma}+e^{y_J}}{2}, \qquad x_2=x_T
\,\frac{e^{-y_\gamma}+e^{-y_J}}{2} \qquad,  
\label{eq:x1x2ycyd} 
\end{equation}
and $x_T=2p_T/\sqrt{s}$. The parton-parton c.m. energy
$\sqrt{\hat{s}}=\sqrt{x_1 x_2 s}$, where $\sqrt{s}$ is the $A+B$
c.m.~energy; the energy of the direct photon in the parton-parton
c.m. system is $P_\gamma^*=E_\gamma^*=\sqrt{\hat{s}}/2$, where  
\begin{equation}
p_T=p_T^*=\frac{\sqrt{\hat{s}}}{2}\sin\theta^* 
\label{eq:shatycyd}
\end{equation}
and $\cos\theta^*=\tanh (y_\gamma-y_J)/2$ is the c.m. angle of the
outgoing $\gamma$ with respect to hadron $A$. 
In Eq.\ref{eq:QCDComptonfull}, $g_A(x_1,Q^2)$
and $g_B(x_2,Q^2)$ are the gluon
structure functions of hadron $A$ and hadron $B$.
At leading order $F_{\rm 2A}(x_1,Q^2)$ and $F_{\rm 2B}(x_2,Q^2)$ 
are structure functions measured in deep inelastic
 scattering (DIS) of $e+A$, given by $F_{\rm 2A}(x,Q^2)=x\sum_a e_a^2\, f^A_a(x,Q^2)$, where $f^A_a(x,Q^2)$ are the distributions in the
number of quarks of type $a$, with electric charge $e_a$ (in units of
the proton charge) in hadron $A$ 
\footnote{The relation $F_2=F_2(DIS)$ 
is true only in leading order pQCD. They can be different in higher 
order pQCD. But those difference is accounted in the theory.}.
In hard $g+q\rightarrow \gamma+q$ scattering in $p$$+$$p$ collisions, 
the struck quark is 8 times more likely to be 
a $u$ quark relative to a $d$ quark. 
For production in nuclei, the ratio is somewhat less according to 
the ratio of the atomic number to the atomic mass.

Beyond leading order, direct photons can be produced either by bremsstrahlung
from any quark line in a 2-to-2 subprocess,
e.g. $g+q\rightarrow g+q+\gamma$, or in a parton shower from
fragmentation that forms a jet. In both these cases the photon is
accompanied by jet fragments so that observing photons isolated from
jets enhances the contribution from the fundamental 2-to-2
subprocesses. Naturally, all these effects must be taken into account
in theoretical calculations of direct photon production in pQCD, and
such calculations~\cite{Aurenche:2006vj} are generally in excellent
agreement with all previous measurements, including those from
PHENIX~\cite{Adler:2006yt}. However, decreasing the uncertainties of
both measurement and theory and extending the range to larger $p_T$ is
desirable. Measurements with and without an isolation criterion allow
more specific comparisons of theoretical models and a better
understanding of photons coming from bremsstrahlung and parton
fragmentation. 

Measurement of direct photons in $p$$+$$p$ collisions provides an important 
baseline for measurements in heavy-ion collisions. Once produced, 
a photon emerges from the reaction almost unaffected since it only interacts 
electromagnetically. Initial state modifications of the distribution functions 
in nuclei can be accessed by measurements in $p+$A or $d+$A collisions. 
Similarly, direct photons provide a reference free from final state effects 
on the outgoing quark, which at LO initially balances the $p_T$ of the direct photon.

In this paper, we report a major update of the direct photon cross section 
measurement in $p$$+$$p$ collisions at $\sqrt{s}=200$ GeV. 
The present data has more than an order of magnitude improved statistics 
than that reported in \cite{Adler:2006yt}. It has extended 
the highest $p_T$ reach of the measurement from 15 GeV/$c$ to 25 GeV/$c$.
We compare the data to pQCD 
calculations and other direct photon data in hadronic collisions.

The paper is organized as follows: Sec. \ref{sec:expset} describes the
experimental setup. Sec. \ref{sec:analysis} describes the analysis method.
Results are in Sec. \ref{sec:results} followed by a
discussion (Sec. \ref{sec:discussion}) and summary
(Sec. \ref{sec:summary}).  The measured invariant cross sections are
tabulated in the Appendix.

\section{Experimental setup}
\label{sec:expset}
Photons were detected in the PHENIX central arm detectors by two 
electromagnetic calorimeter (EMCal) arms
(West, East) each covering $\pi/2$ rad in azimuthal angle ($\phi$) 
and $|\eta|<0.35$ in pseudorapidity. Each arm is
divided into 4 sectors in azimuth. All 4 sectors in the West arm are 
lead scintillator
sampling detectors (PbSc). In the East arm, 2 sectors are PbSc and 
2 sectors are lead glass \v{C}erenkov detectors (PbGl).
The sectors are composed of independent towers with granularities 
of $\Delta \eta \times \Delta \phi =
~0.011\times0.011$ and $ ~0.008\times0.008$ for the PbSc and the PbGl,
respectively. 
A tower contains $\sim\!80$\% of the photon energy hitting the center of the tower.
During data taking, the relative gain of the detectors
was monitored using a light pulser calibration system.  The absolute
energy calibration was based on the known minimum-ionizing energy
peak of charged tracks, energy-momentum matching of identified
electron tracks and the measured value of the $\pi^0\rightarrow\gamma\gamma$
invariant mass. The linearity of the energy response was obtained
from beam tests and the dependence of the measured $\pi^0$ mass on its momentum.  
The energy resolution was determined using the width of the $\pi^0$ peak,
and was $\sigma_E/E=8.1\%/\sqrt{E}(\rm{GeV})\oplus 5.7\%$.
The systematic error on the absolute energy scale is less than 1.5\%. 
The time of flight (ToF) as measured by the EMCal with a resolution of
better than 1 ns was used to reduce the cosmic ray background.

The dynamic range of the electronics is saturated for the highest
energy clusters ($\sim 25$ GeV) measured in this analysis. 
The size of this effect is
estimated using a convolution of the maximum energy limit of each
tower (26 GeV typical) and the fraction of energy deposited in the 
central tower of a electromagnetic shower cluster. This effect was 
found negligible compared to the statistical uncertainty for the 
very high energy photons that are affected.

The drift chambers (DC) and the innermost layer of the pad chambers
(PC1) provide charged track information and were used to veto charged
hadron clusters in the EMCal.
Hits in the beam-beam counters (BBCs) positioned at pseudorapidities 
$3.1\!<\!|\eta|\!<\!3.9$ were used 
to measure a collision vertex from the time difference between hits 
in both BBCs, and to monitor the luminosity.
The PHENIX detector is described in detail elsewhere \cite{Adcox:2003zm}. 

\section{Analysis}
\label{sec:analysis}

\subsection{Event selection}

The results in this paper are based on the data sample taken during 
the 2006 RHIC run.  A high $p_T$ photon sample was collected with 
an EMCal trigger in which the analog sum of signals from a $4\times 
4$ adjacent set of EMCal towers was greater than a nominal energy 
threshold of 2 GeV, which was in coincidence with the minimum-bias 
trigger, corresponding to $\sim\!8.0$ pb$^{-1}$ integrated 
luminosity. The integrated luminosity was determined from the rate 
of a minimum-bias trigger that required hits in the BBCs and a 
collision vertex within 30 cm of the nominal center of the 
interation region. At $\sqrt{s}=200$ GeV this trigger selects 23.0 
mb of the inelastic $p+p$ cross section. This was measured with 
9.7\% uncertainty using a Vernier-scan 
technique~\cite{Adler:2003pb,Adare:2008qb}. This corresponds to 
about 55\% of the inelastic $p$$+$$p$ cross section.  For the BBC 
trigger rate of 250 kHz typical of the sample used in this analysis 
in the crossing rate of 9.4 MHz, the effect of multiple collisions 
per bunch crossing was $<2$\% and is neglected.

\subsection{Photon selection}

Photon candidates were reconstructed from EMCal clusters within an
EMCal fiducial volume defined to exclude the edge areas $\Delta
\eta$=0.10 and $\Delta \phi$=0.10 wide, resulting in a fiducial area
of $|\eta_{f}|<0.25$ in pseudorapidity and $\Delta
\phi_{f}=\pi/2-0.2$ in azimuth for each of the two arms.  Areas
outside the fiducial volume were included when searching for a partner
candidate for $\pi^0 \rightarrow \gamma \gamma$ decays to suppress
$\pi^0$ background, and to measure activity around direct photon
candidates (an isolation cut, section \ref{sec:iso_method}) to suppress 
bremsstrahlung and fragmentation photons in the reconstructed photon 
sample.

Photon candidates were required to have $p_T>5$ GeV/$c$ and a ToF 
measured by the EMCal to be within $\pm5$ ns from the expected 
arrival time for the photons originating at the vertex.  This 
requirement reduced the background from cosmic rays by an order of 
magnitude. The remaining contributions are estimated from the ToF 
distribution and corrected. The magnitude of this correction was 
negligible in the region of $p_T<\sim\! 15$ GeV/$c$ and larger for 
higher $p_T$ photons as the rate goes down ($\sim\! 8$\% 
contamination at $p_T=25$ GeV/$c$). The background from charged 
tracks was suppressed by requiring that the EMCal cluster shape be 
consistent with a single electromagnetic shower and that no charged 
track points to to the cluster. The shower shape cut efficiency for 
photons, evaluated using reconstructed $\pi^0$ decay photons, was 
0.98 over the relevant $p_T$ range.  Most of the conversion 
$e^+e^-$ pairs in the $\sim10$\% of radiation length of material 
between the DC and the EMCal are reconstructed as single photons 
because of the minimal magnetic field in this region. Of course, no 
charged track would point to the cluster since the conversion would 
have happened after the DC. An additional 1\% loss was attributed 
to these photon conversions from a {\sc geant} \cite{GEANT:W5013} 
simulation with a reasonable input $p_T$ distribution. The 
contribution of the other hadronic background (neutrons, $K_L$, 
albedo from magnet poles and other material) was studied with a 
detailed {\sc geant} Monte Carlo simulation and found to be less 
than 1\% of the photon sample.

The fine granularity of the PHENIX EMCal
resolves the two photons from $\pi^0 \rightarrow \gamma \gamma$ decays
up to $\pi^0$ $p_T$ of 12 GeV/$c$ (17 GeV/$c$) in the PbSc (PbGl).  
A 50 \% merging probability corresponds to $\pi^0$ $p_T$ of 17 GeV/$c$ 
(25 GeV/$c$). In the $p_T$ range presented in this paper (up to 25
GeV/$c$), merged photons can be separated from single photon showers
in the EMCal and rejected using shower shape measurements with an 
efficiency $>90\%$.

\subsection{Direct photon signal extraction}
\begin{figure}[thb]
\includegraphics[width=1.0\linewidth]{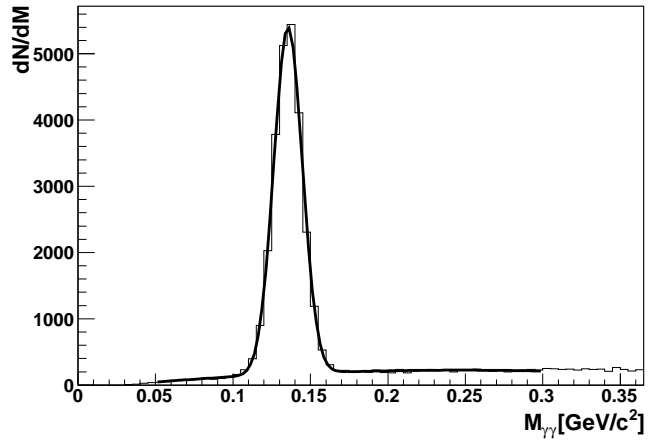}
\caption{Two-photon invariant mass distribution in the West arm where one 
of the photons has $5<p_T<5.5$ GeV/$c$.}
\label{fig:mgg_west5}
\end{figure}

In the obtained photon sample the majority of the background for
direct photon measurements comes from decays of hadrons, primarily
$\pi^0 \rightarrow \gamma \gamma$ ($\sim\! 80\%$) and $\eta \rightarrow
\gamma \gamma$ ($\sim\! 15\%$).  The contribution from $\pi^0$ decays
was evaluated by a $\pi^0$-tagging method \footnote{In the previous 
measurement~\cite{Adler:2006yt}, we introduced a $\pi^0$-tagging 
method and a cocktail subtraction method. Both use statistical subtraction,
but the $\pi^0$-tagging method uses the photon $p_T$ on which 
it is easier to apply an isolation cut.}. 
In this approach the direct
photon candidate was paired with each of the other photons in an event
(a partner photon) to calculate the two-photon invariant mass $M_{\gamma
  \gamma}$, which was required to be in the range from 105 to 165 MeV
corresponding to $\pm 3\sigma$ around the $\pi^0$ peak (see
Fig. \ref{fig:mgg_west5}).  Both photons were required to have a
minimum energy $E_{\rm min}=0.5$~GeV.

\begin{figure}[thb]
\includegraphics[width=1.0\linewidth]{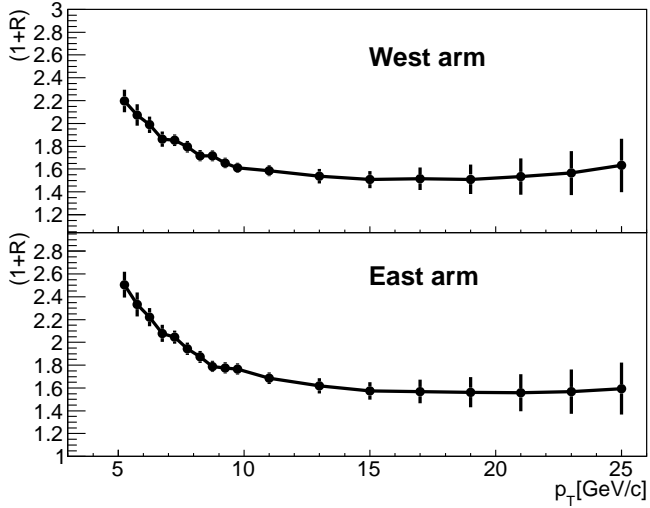}
\caption{The correction factor to be applied to the number of tagged
  photons for the total contribution from $\pi^0$ ((1+R) in
  Eq.\ref{eqn:subtraction}) from a single particle Monte Carlo
  simulation. The error bar shows the systematic uncertainty.}
\label{fig:pi0miss}
\end{figure}

The combinatorial background under the $\pi^0$ peak was evaluated and
then subtracted by fitting the two-photon invariant mass distribution
outside the peak region. The corrections for the underestimation 
due to photon conversions and $\pi^0$ Dalitz decays were applied as 
a part of the partner photon efficiency.  
To avoid
acceptance losses for $\pi^0$ reconstruction, the edge areas of the
EMCal outside the fiducial region were included for partner photon
selection as long as the primary photon was in the fiducial region.
The minimum photon energy cut, 
the EMCal geometry, and inactive areas
led to an underestimate of the $\pi^0$ decay
photon yields. A correction for this was calculated using a single
particle Monte Carlo simulation, which included the EMCal geometry,
the configuration of dead areas, resolution, and the $\pi^0$ spectrum
shape from earlier measurements \cite{Adare:2008qb}.  Figure
\ref{fig:pi0miss} shows a $p_T$ dependent multiplicative correction,
denoted as $1+R$, to the tagged $\pi^0$ photon sample.  1+R drops 
with increasing $p_T$ when
going from 5 GeV/$c$ to 15 GeV/$c$ due to the decreasing influence of the
$E_{\rm min}$ cut. As the $p_T$ increases further the correction stops 
decreasing due to an increasing merging probability. The correction for
asymmetric decays is more affected by the $E_{\rm min}$ cut, 
since symmetric decays start merging and are rejected from the photon 
sample by the shower shape cut.
Completely merged photons from high $p_T$ $\pi^0$ decays may
resemble single photons in the EMCal; this residual 
contribution is corrected later. 

The contribution to the photon sample from hadronic decays other than
from $\pi^0$s was estimated relative to the $\pi^0$ decay contribution
based on the $\eta / \pi^0$ \cite{Adare:2010cy}, and $\omega / \pi^0$
and $\eta' / \pi^0$ \cite{Adare:2010fe} ratios from our measurements
assuming $m_T$ scaling (tested in \cite{Adare:2010fe}). 
We denote the ratio of photons from these 
decays to the photons from $\pi^0$ decays as $A$.  
At lower $p_T$ ($p_T\sim 5$ GeV/$c$) this ratio, $A$,  has a weak 
$p_T$ dependence and approaches 0.235 as $p_T$ increases. 
Since the contribution of photons from $\pi^0$ to the
background decreases for $p_T>10$ GeV/$c$ due to $\pi^0$ photons merging, 
the value, $A$, starts to rise linearly at around $p_T=12$ GeV/$c$, and 
at the highest $p_T$ point at 25 GeV/$c$ it is 1.4 (0.94) 
for West (East) arm.

The yield of direct photons $N_{\rm dir}$ was obtained from the inclusive
photon yield $N_{\rm incl}$ as follows:
\begin{equation}
N_{\rm dir} = N_{\rm incl} - (1+A)(1+R)N_{\pi^0}, 
\label{eqn:subtraction}
\end{equation}
where $N_{\pi^0}$ is the contribution from $\pi^0$s evaluated with a
tagging process. In this notation, $(1+R) \cdot N_{\pi^0}$ represents 
the total contribution from unmerged $\pi^0 \rightarrow \gamma \gamma$ 
decays, and $A \cdot (1+R) \cdot N_{\pi^0}$ is the contribution 
from other hadronic decays. 

\begin{figure}[thb]
\includegraphics[width=1.0\linewidth]{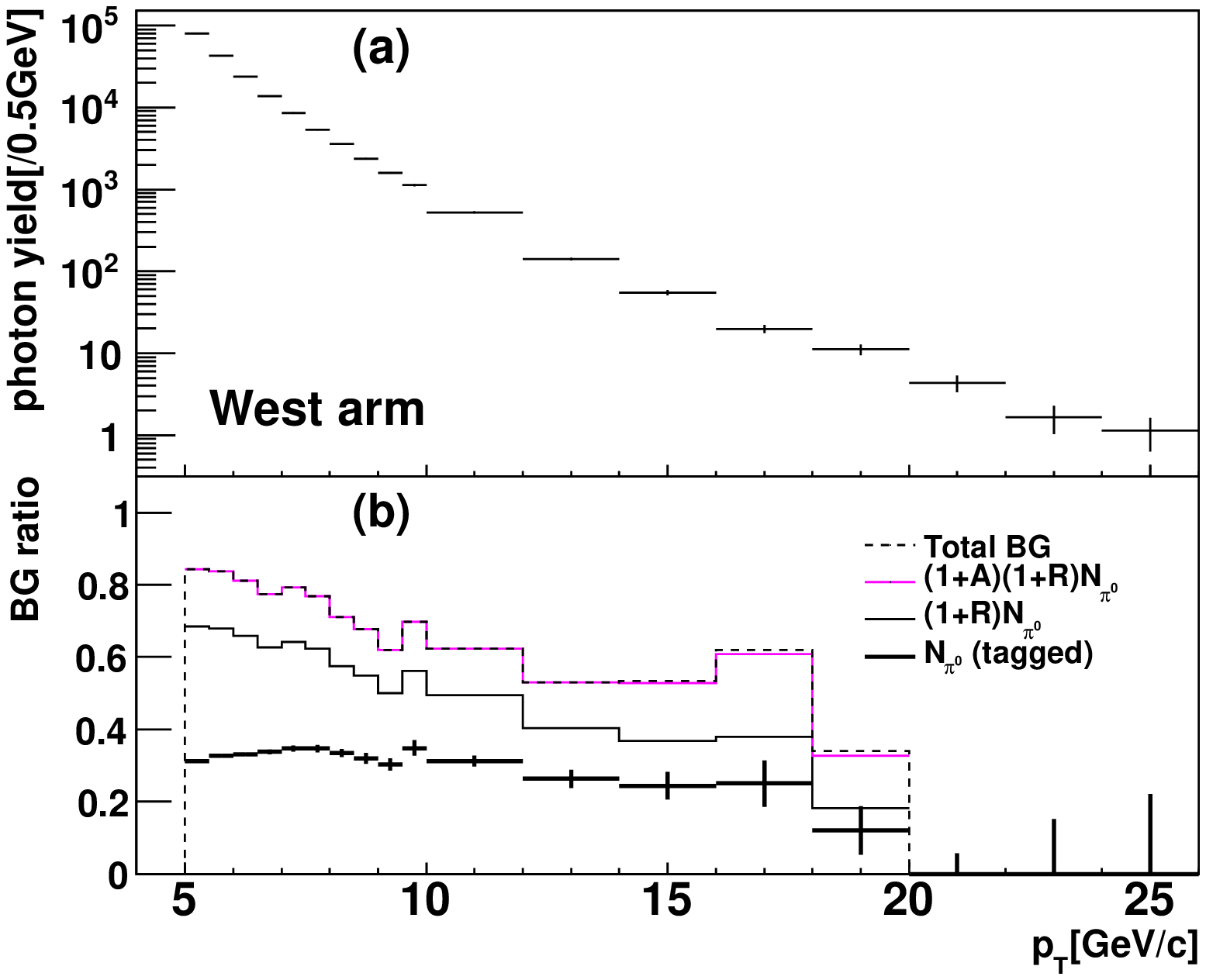}
\caption{(color online) In the West arm, (a) number of photon 
  clusters and (b) Background-photon fraction.  
From bottom to top, tagged $\pi^0$ ($N_{\pi^0}$), total
  photons from unmerged $\pi^0$ ($(1+R)N_{\pi^0}$), all hadronic decay
  ($(1+A)(1+R)N_{\pi^0}$), and the total background including an
  estimate of completely merged clusters. In the highest 3 bins where 
  no $\pi^0$-tagged photons were found, only the uncertainties of 
  $N_{\pi^0}$ are shown.}
\label{fig:compsub_w}

\includegraphics[width=1.0\linewidth]{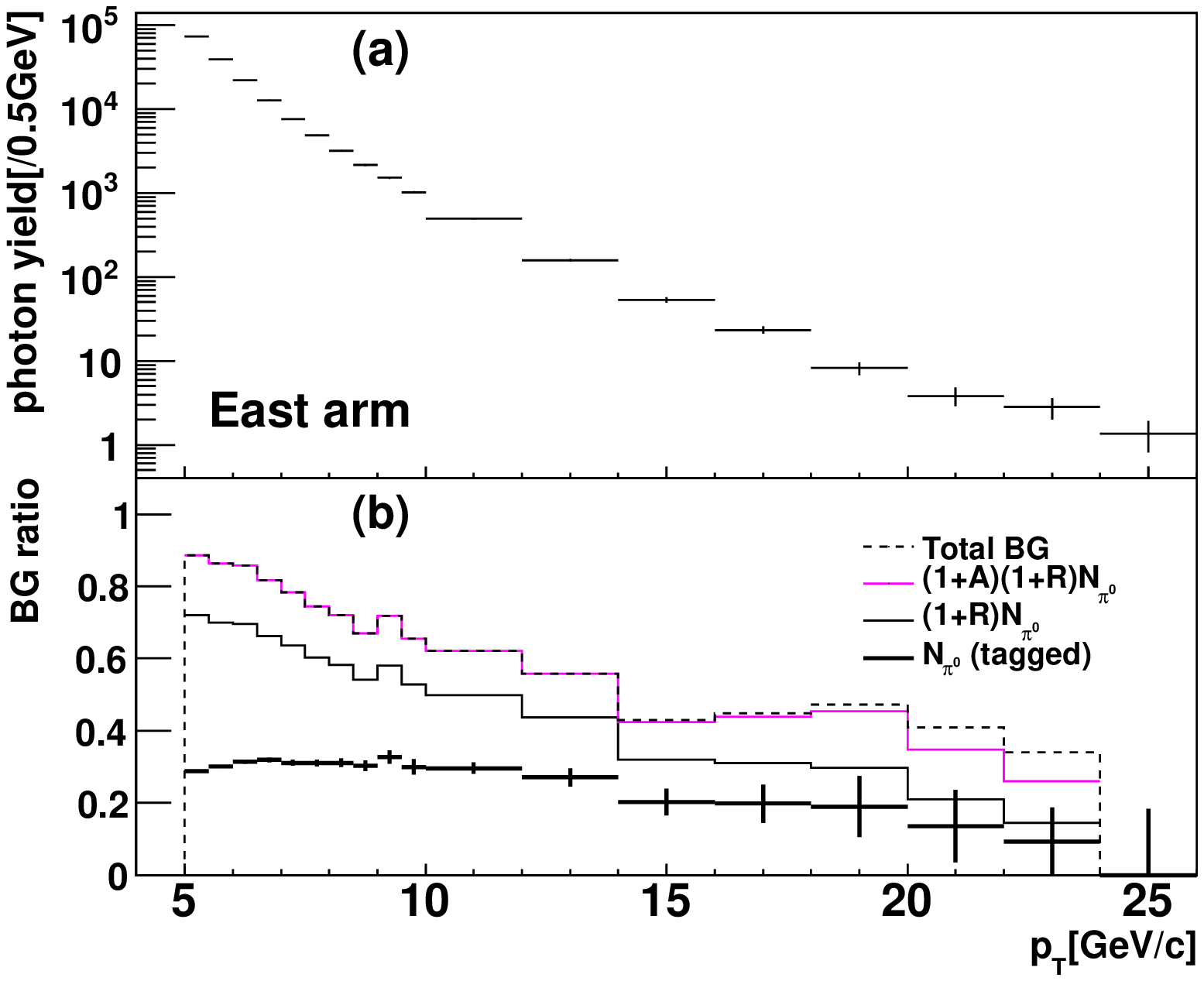}
\caption{(color online) The same as Fig. \ref{fig:compsub_w} but for the East arm.}
\label{fig:compsub_e}
\end{figure}

Figures \ref{fig:compsub_w} and \ref{fig:compsub_e} show different
contributions to the inclusive photon spectrum separately for West
and East spectrometers. In the highest $p_T$ bins, where no
$\pi^0$-tagged photons were found (due to low statistics and high
merging probability), $N_{\pi^0}$ was set to $0^{+1}_{-0}$ 
to safely cover the other hadronic-decay channels.

\subsection{Direct Photon Cross Section}

Based on the extracted direct photon yields $N_{\rm dir}$ in each $\Delta p_T$ wide
bin in transverse momentum, the invariant cross section of direct photon
production was calculated as follows:

\begin{equation}
E\frac{d^3\sigma}{dp^3}=\frac{1}{\mathcal{L}}\frac{1}{2\pi p_T}\frac{N_{\rm dir}}{\Delta p_T\Delta y}\frac{1}{\epsilon}\frac{1}{\epsilon_{bias}},
\label{eqn:crosssection}
\end{equation}
where $\mathcal{L}$ is the integrated luminosity for the analyzed data sample,
$\Delta y$ is the rapidity range, $\epsilon$ corrects for
the acceptance including the trigger live area, photon reconstruction efficiency, 
trigger efficiency and $p_T$ smearing due to EMCal energy resolution.  
$\epsilon_{bias}$
corrects for the finite efficiency of BBCs to trigger on high $p_T$
events.  The latter was measured from the ratio of $\pi^0$ yields
obtained using the EMCal based high $p_T$ photon trigger with and
without BBC trigger requirements, and was found to be $0.78 \pm 0.02$,
independent of $p_T$.

\begin{figure}[thb]
\includegraphics[width=1.0\linewidth]{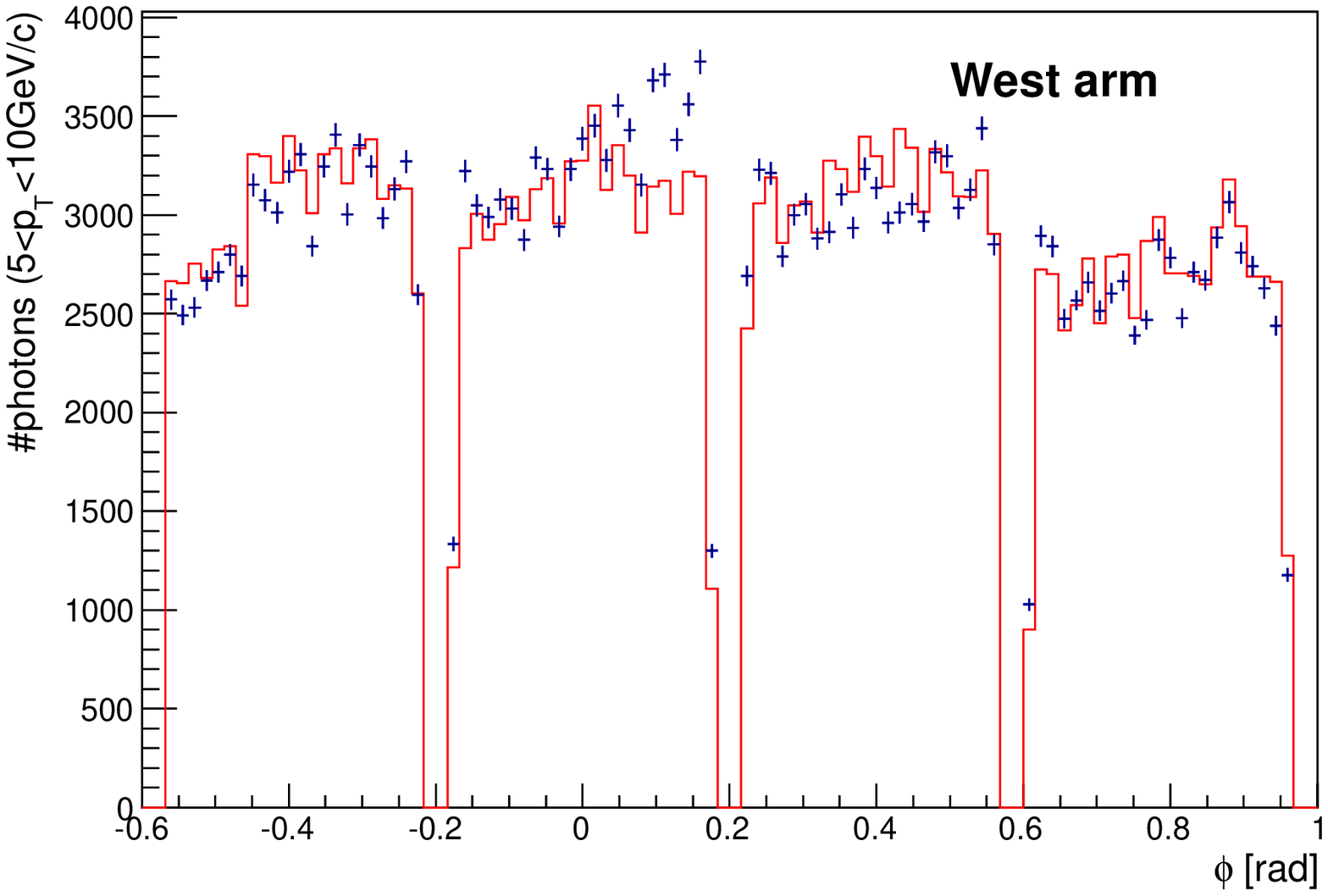}
\caption{(color online) Number of photons ($5\!<\!p_T\!<\!10$ GeV/$c$) 
  as a function of the azimuthal angle (West arm). The histogram shows
  the MC result normalized by the total count.}
\label{fig:cysecw}

\includegraphics[width=1.0\linewidth]{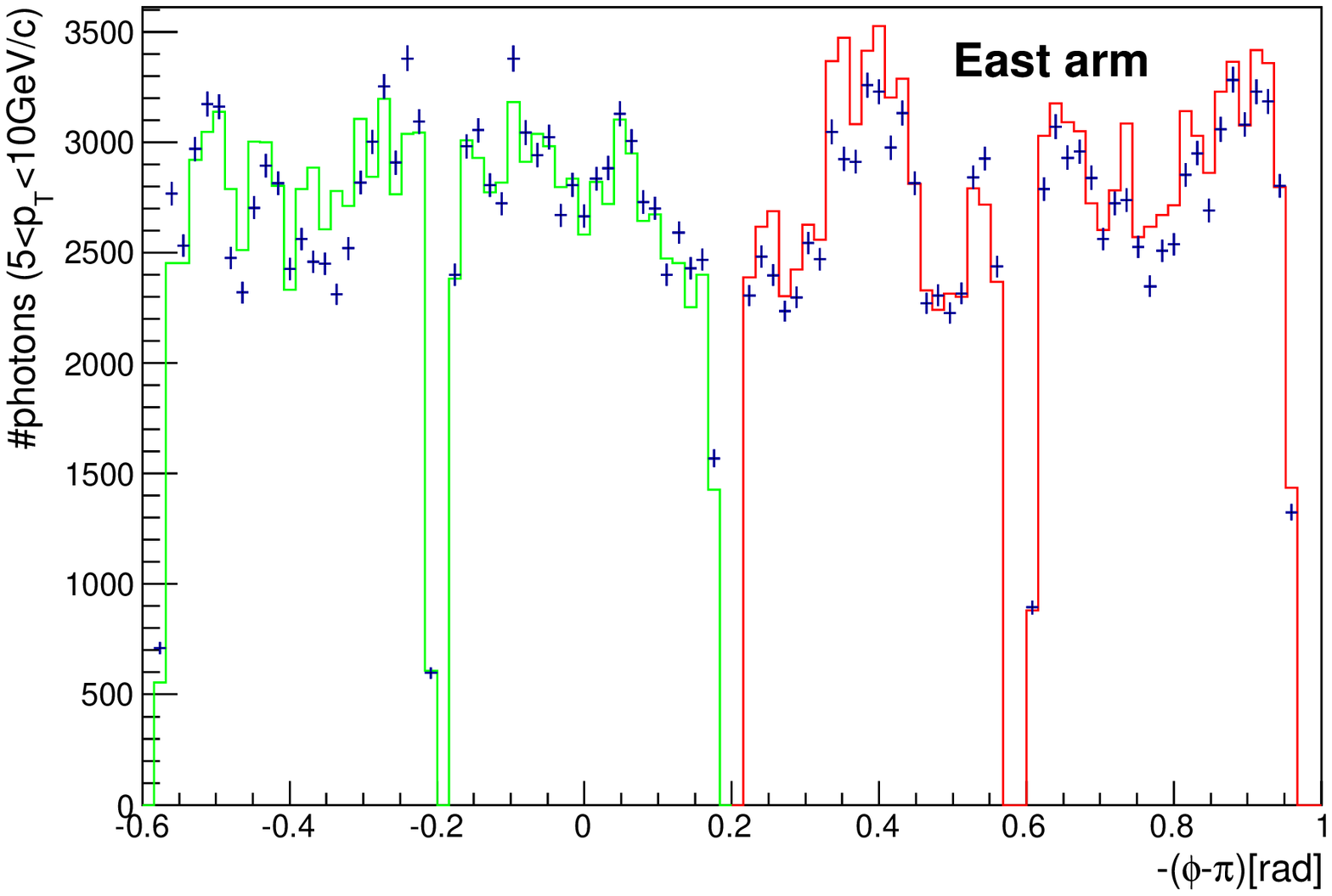}
\caption{(color online) Number of photons ($5\!<\!p_T\!<\!10$ GeV/$c$) 
  as a function of the azimuthal angle (East arm). The lower (upper)
  half corresponds to PbGl (PbSc) sectors. The histogram shows the MC
  result normalized by the total count (PbSc and PbGl sectors are done
  separately.)}
\label{fig:cysece}
\end{figure}

A single particle Monte Carlo simulation, which included the
configuration of detector active areas and resolutions, was used to
evaluate the corrections for the acceptance and $p_T$ smearing. 
The small differences between the distribution of azimuthal position of
reconstructed photons between data and simulation
(Fig. \ref{fig:cysecw} and \ref{fig:cysece}) served to estimate the
systematic uncertainty.  
The effects of $p_{T}$ smearing were determined by varying the 
input $p_{T}$ spectrum of the simulation.
The same simulation framework was used to evaluate the propagation of the
1.5\% scale uncertainty in the EMCal energy measurements to the final
direct photon spectrum.

The terms from Eq. \ref{eqn:subtraction} used to calculate $N_{\rm dir}$ contribute
to its uncertainty as follows (as tabulated in Table \ref{table:syserror}):
\begin{eqnarray}
\nonumber \frac{\delta N_{\rm dir}}{N_{\rm dir}}=W \cdot (\frac{\delta N_{\rm incl}}{N_{\rm incl}}) 
     \oplus (W-1)\cdot(\frac{\delta (1+A)}{(1+A)}) \\
     \oplus (W-1)\cdot(\frac{\delta(1+R)}{(1+R)})
     \oplus (W-1)\cdot(\frac{\delta N_{\pi^0}}{N_{\pi^0}}),
\label{eqn:subtraction_sys}
\end{eqnarray}
where $W$ is defined as $N_{\rm incl}/N_{\rm dir}$.  At low $p_T$ with high
backgrounds ($W>>1$), the contributing uncertainties on $N_{\rm dir}$ are
amplified by a factor $W$ or $W-1$.  In this $p_T$ range the dominant
uncertainty comes from the correction for untagged photons ($(1+R)$)
from $\pi^0$-decay due to its sensitivity to the minimal energy cut,
$E_{\rm min}$.  In the higher $p_T$ bins the biggest systematic
uncertainty of the component is in the merging effect (which is included 
in the (1+R) term), since most of $\pi^0$s are merged in the EMCal in this $p_T$
region.  However this effect on the direct photon is small, 
suppressed by a factor $(W-1)<<1$, because of the small background 
fraction and approaches zero at highest $p_T$.

\subsection{Study of the effect of isolation cut}
\label{sec:iso_method}
In this section we investigate the effect of an isolation cut on 
direct photons 
to determine the fraction coming from Compton and
annihilation processes, which are expected to be isolated from jet
activity.  The energy around the photon candidate in a cone of radius
$r=\sqrt{(\delta \eta)^2 + (\delta \phi)^2}=0.5$ was required to be
less than 10\% of the photon energy, in order to pass the isolation
cut. The cone size is determined by the hadron correlation in a jet 
measurement (e.g. Fig. 6 in \cite{Adare:2010yw}).
The total energy in the cone was constructed by summing the energy of
electromagnetic clusters in the EMCal and the momentum of charged
tracks in the tracking system.  To be counted as part of the cone
energy, the minimum EMCal cluster energy was set at 0.15 GeV
and the minimum track momentum at 0.2 GeV/$c$, close to the lower
limit for charged particle reconstruction in PHENIX.  To avoid inclusion of
misidentified tracks, which, due to decays or photon conversions may
mimic high $p_T$ tracks, the maximum momentum for the tracks
to participate in the cone energy calculation was set to 15 GeV/$c$.  

Aside from direct photons, the isolated photon sample ($N_{\rm incl}^{\rm iso}$)
includes background photons from $\pi^0$ and other hadron decays.
For $\pi^0$ decays we consider photons that have a partner photon 
reconstructed in the EMCal acceptance ($n_{\pi^0}^{\rm iso}$), and those  
which satisfy the isolation criteria if the partner 
photon is masked out ($N_{\pi^0}^{\rm iso}$). 
More $\pi^0$ photons pass the isolation cut without
the partner photon energy, so $n_{\pi^0}^{\rm iso}$ is a 
subgroup of $N_{\pi^0}^{\rm iso}$.
The latter was used to estimate the isolated photons from $\pi^0$ 
with missing partner due to the energy threshold $E_{\rm min}$ or
EMCal masked areas, by multiplying by the same missing photon fraction 
$R$ introduced in Eq. \ref{eqn:subtraction}. 

To estimate the contribution of other hadron decays, the isolated
photon candidate from $\pi^0$ ($N_{\pi^0}^{\rm iso}$) is scaled. With this
procedure, the isolation cut efficiency from jet fragments is taken
into account, however there is an additional rejection due to its own
partner photon. A single particle MC for $\eta$s was used to include
this effect. In the case of $\eta$s, for the lowest $p_T$ sample, the
partner photon can be out of the EMCal acceptance because of a large
opening angle, thereby reducing the rejection power. As it goes to
high $p_T$, the rejection power due to the partner energy becomes
constant.

Similar to Eq.\ref{eqn:subtraction}, 
the isolated direct photon yield ($N_{\rm dir}^{\rm iso}$) was calculated
using:
\begin{equation}
N^{\rm iso}_{\rm dir}=N^{\rm iso}_{\rm incl}-(n^{\rm iso}_{\pi^0}+N^{\rm iso}_{\pi^0}R)-A^{\rm iso}(1+R)N^{\rm iso}_{\pi^0}
\label{eqn:isolation}
\end{equation}

\begin{figure}[thb]
\includegraphics[width=1.0\linewidth]{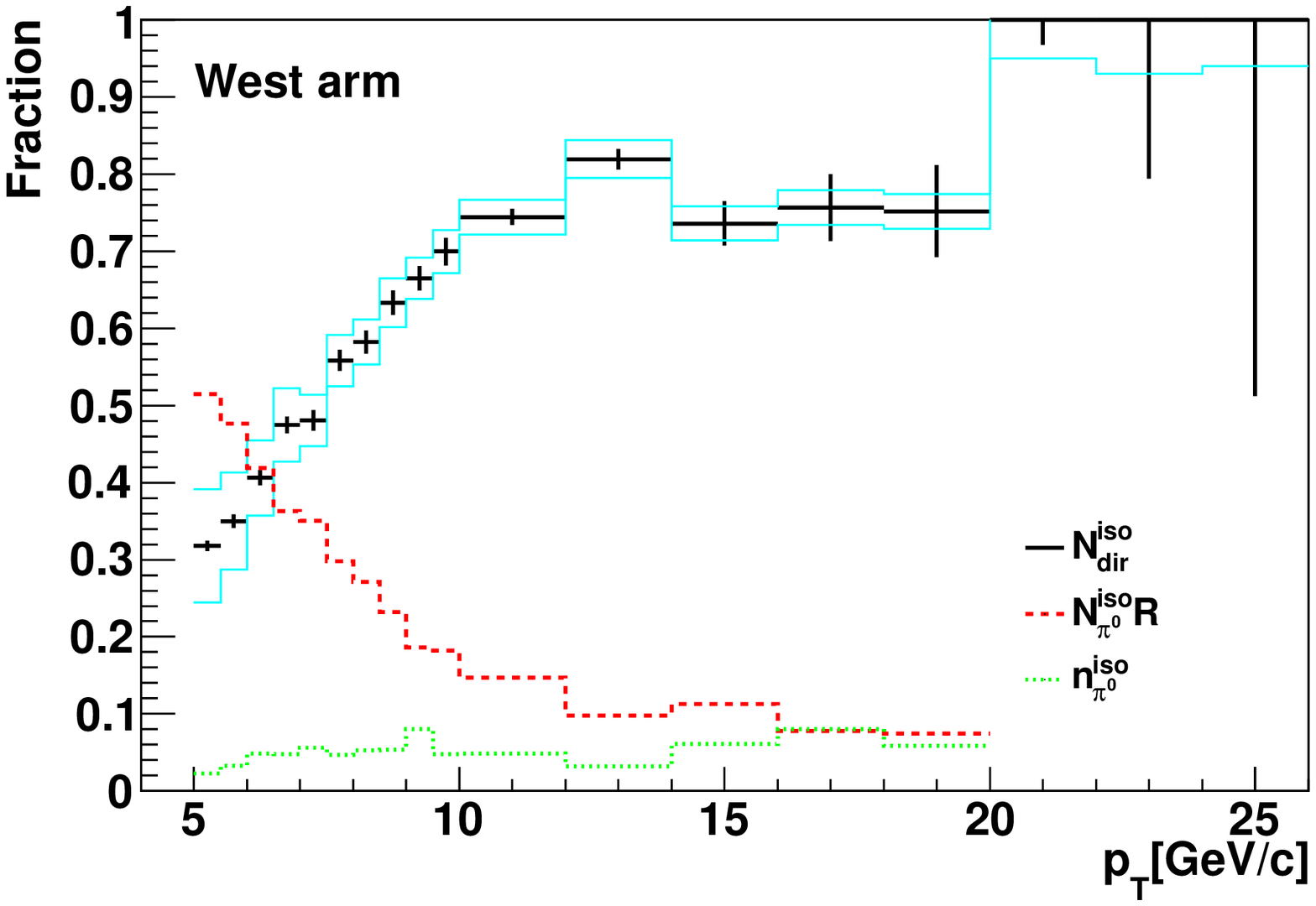}
\caption{(color online) Components of isolated photon clusters (West arm). Solid line (with error): isolated direct photon signal ($N_{\rm dir}^{\rm iso}$), 
 Dashed line : photons from $\pi^0$ with missing partner ($N^{\rm iso}_{\pi^0}R$), Dotted line : Photons tagged as 
$\pi^0$ ($n^{\rm iso}_{\pi^0}$).}
\label{fig:compiso_w}

\includegraphics[width=1.0\linewidth]{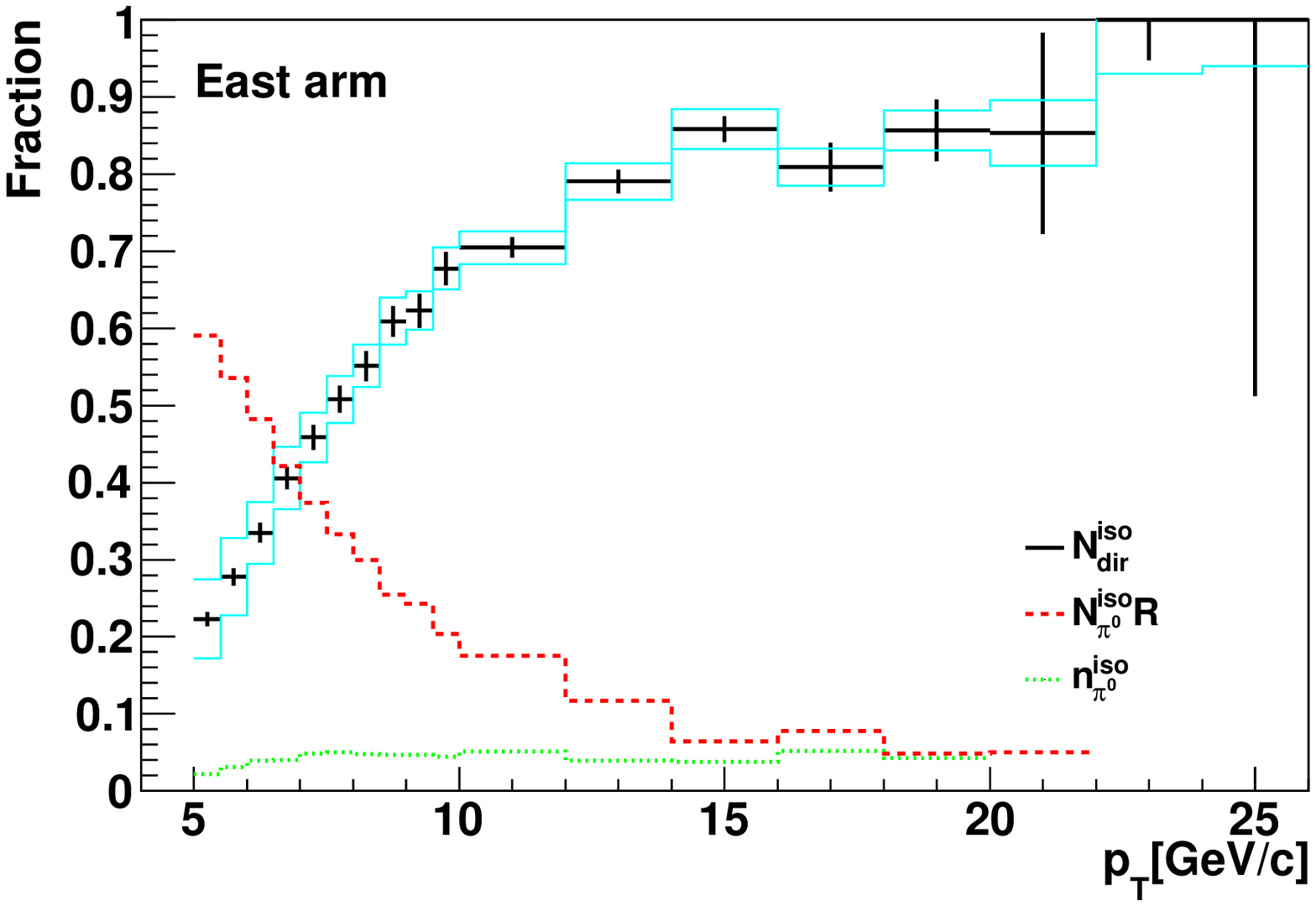}
\caption{(color online) Components of isolated photon clusters (East arm)}
\label{fig:compiso_e}
\end{figure}

In addition to $A$ in Eq.\ref{eqn:subtraction}, $A^{\rm iso}$ includes
the isolation cut effect of the hadron's own partner photon 
as described above. 
Different contributors to the isolated photon sample $N_{\rm incl}^{\rm iso}$
are shown in Fig. \ref{fig:compiso_w} and \ref{fig:compiso_e}.

Uncertainties are propagated according to:
\begin{eqnarray}
\label{eqn:isolation_sys}
\nonumber \frac{\delta N^{\rm iso}_{\rm dir}}{N^{\rm iso}_{\rm dir}} & = & W_0\frac{\delta N^{\rm iso}}{N^{\rm iso}_{\rm incl}}\oplus W_1\frac{\delta (1+A^{\rm iso})}{(1+A^{\rm iso})} \\ 
\nonumber &   & \oplus W_1\frac{\delta (1+R)}{(1+R)}\oplus W_2\frac{\delta n_{\pi^0}^{\rm iso}}{n_{\pi^0}^{\rm iso}}\oplus W_3\frac{\delta N_{\pi^0}^{\rm iso}}{N_{\pi^0}^{\rm iso}}\\
\nonumber W_0 & = &\frac{N^{\rm iso}_{\rm incl}}{N^{\rm iso}_{\rm dir}}, \\
\nonumber W_1 & = &\frac{(1+A^{\rm iso})(1+R)N_{\pi^0}^{\rm iso}}{N^{\rm iso}_{\rm dir}}, \\
\nonumber W_2 & = &\frac{n_{\pi^0}^{\rm iso}}{N^{\rm iso}_{\rm dir}}, and \\
\nonumber W_3 & = &\frac{((1+A^{\rm iso})(1+R)-1)N_{\pi^0}^{\rm iso}}{N^{\rm iso}_{\rm dir}}. \\
\end{eqnarray}

Smoothed functions from fits to the data of the $W_0$, $W_1$, $W_2$,
and $W_3$ parameters were used. The overall trend is the same as in
the case of inclusive photon measurement. However at low $p_T$ the
systematic uncertainties are smaller due to smaller contribution from
hadronic decay photons in the isolated photon sample.  Appendix
Table \ref{table:isosyserror} summarizes the systematic uncertainties for isolated
direct photon measurements.

\begin{figure}[th]
\includegraphics[width=1.0\linewidth]{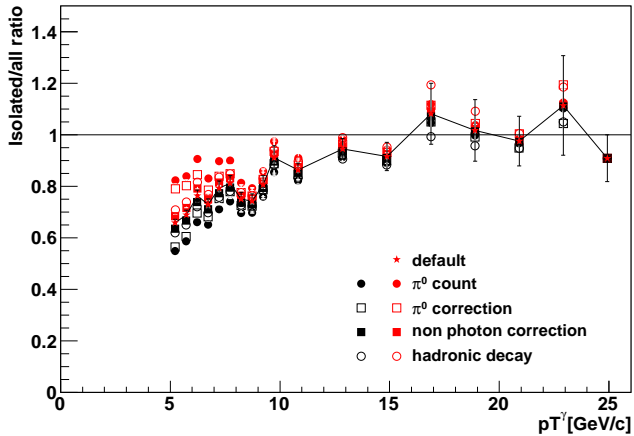}
\caption{(color online) 
Effects in the ratio of isolated-direct photons for 
(black points) $-1\sigma$ and
(red points) $+1\sigma$ 
of systematic uncertainty change: 
(closed circles) $\pi^0$ counts, 
(open squares) untagged $\pi^0$ corrections, 
(closed squares) nonphoton contributions, 
(open circles) hadronic components other than $\pi^0$.}
\label{fig:isosubparscan}
\end{figure}

\subsection{Isolation over inclusive-direct photon ratio}
By taking the ratio of isolated direct photons to the inclusive-direct
photons, some uncertainties such as photon efficiency and the
luminosity measurement cancel.  To estimate the remaining
systematic uncertainty, each component contributing to the yield
measurements was moved up and down by $\pm 1\sigma$ of its systematic
uncertainty.  Figure \ref{fig:isosubparscan} shows the variation due
to this change.  The values are tabulated in Appendix
Table \ref{table:syserrorisoratio}.  The total systematic uncertainty was
taken as the quadratic sum of these variations.

To evaluate the rejection power of the isolation cut on jet
fragmentation, the same ratio was calculated for photons from $\pi^0$
($=N^{\rm iso}_{\pi^0}/N_{\pi^0}$).  Here the systematic uncertainty is
from the correction of combinatorial background.

\begin{figure}[th]
\includegraphics[width=1.0\linewidth]{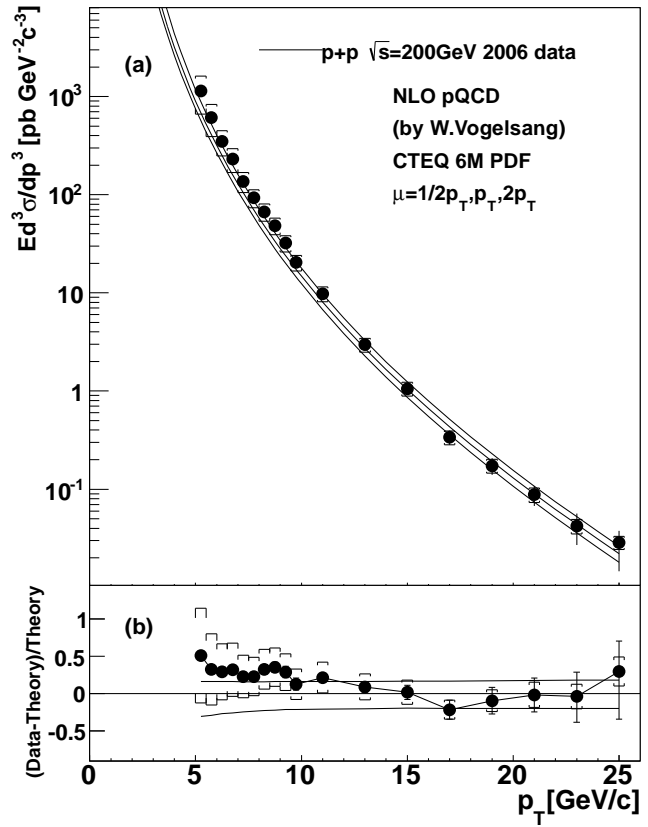}
\caption{(a) Inclusive direct photon spectra compared with an
  NLO pQCD calculation for three different renormalization and 
  factorization scales as
  mentioned in the text (The same calculation was used in \cite{Adler:2006yt}). 
  The error bars denote point-by-point
  uncertainties, the error brackets show $p_T$ correlated
  uncertainties. (b) Comparison of the data and the pQCD
  calculation.}
\label{fig:spectrasub}
\end{figure}
 
\section{Results}
\label{sec:results}
Figure \ref{fig:spectrasub} shows the inclusive-direct photon cross
section where the bin width correction in the data are applied to the
vertical direction.  The data are compared to a NLO pQCD calculation
\cite{Gordon:1993qc,Gordon:1994ut,Aurenche:1983ws,Aurenche:1987fs,Baer:1990ra,Baer:1989xj}
using the CTEQ 6M parton distribution functions \cite{Pumplin:2002vw}
and the BFGII parton to photon fragmentation function (FF)
\cite{Bourhis:1997yu} for three different renormalization and
factorization scales, from bottom to top, $\mu$=2$p_T$, $p_T$ and
$p_T/2$.  The deviation of
the data from the calculation is shown in the bottom panel.
The present data are consistent with that we previously published in 
\cite{Adler:2006yt}.
The highest $p_T$ reach of the data is expanded from 15 GeV/$c$ to 25 GeV/$c$.
For the $p_T$ range ($8<p_T<25$ GeV/$c$) a power law fit to the data gives
$n= 7.08\pm0.09(stat)\pm0.1(syst)$ with $\chi^2/NDF=8.3/10$. 
In the fit, all systematic uncertainties are treated as correlated.

\begin{figure}[thb]
\includegraphics[width=1.0\linewidth]{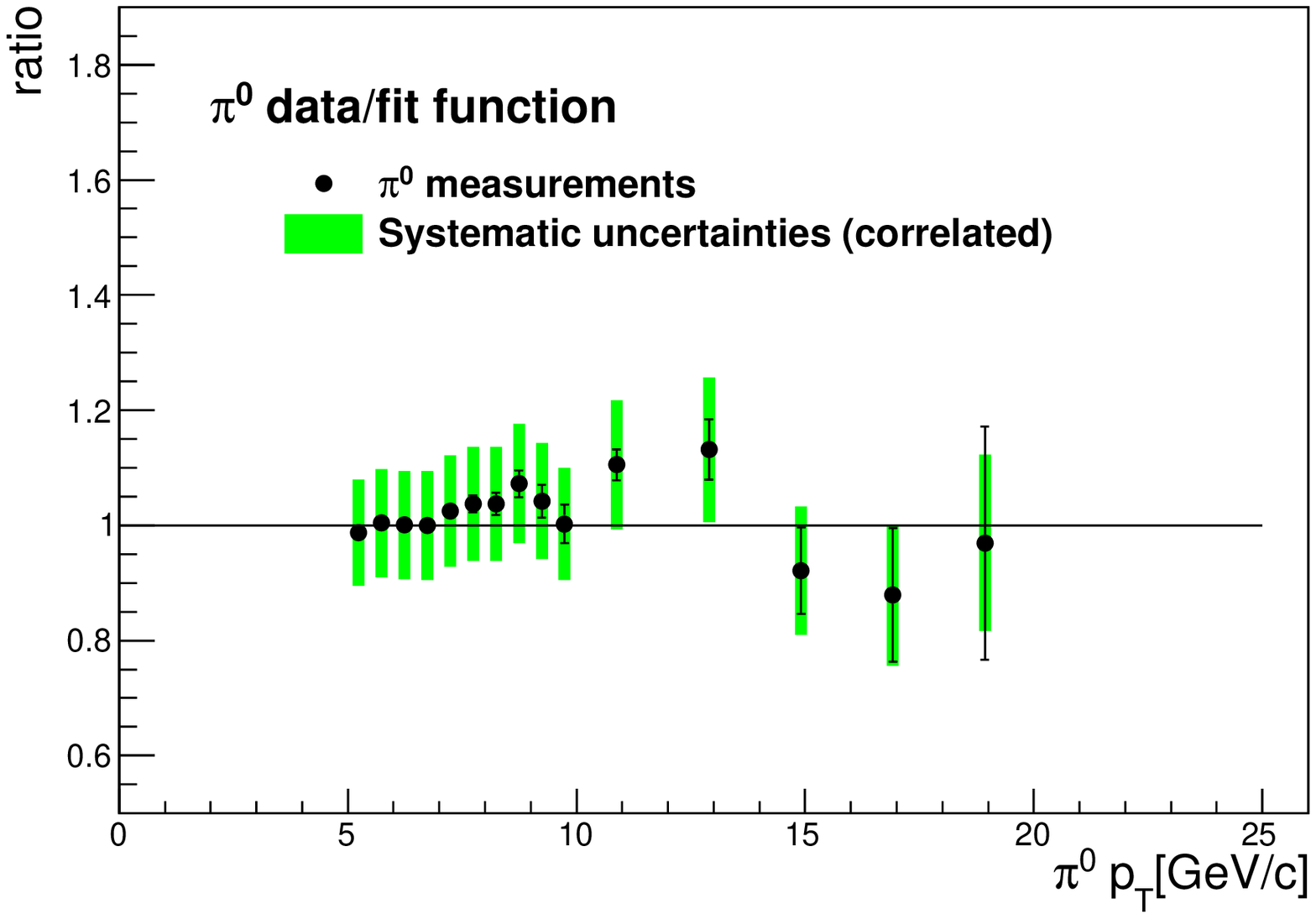}
\caption{(color online) The ratio of $\pi^0$ spectrum \cite{Adare:2007dg} to the power law fit for the
  range $5<p_T<20$ GeV/$c$.  The systematic uncertainty shown by the box
  does not include the overall normalization uncertainty.}
\label{fig:crfit_pow1777_822}

\includegraphics[width=1.0\linewidth]{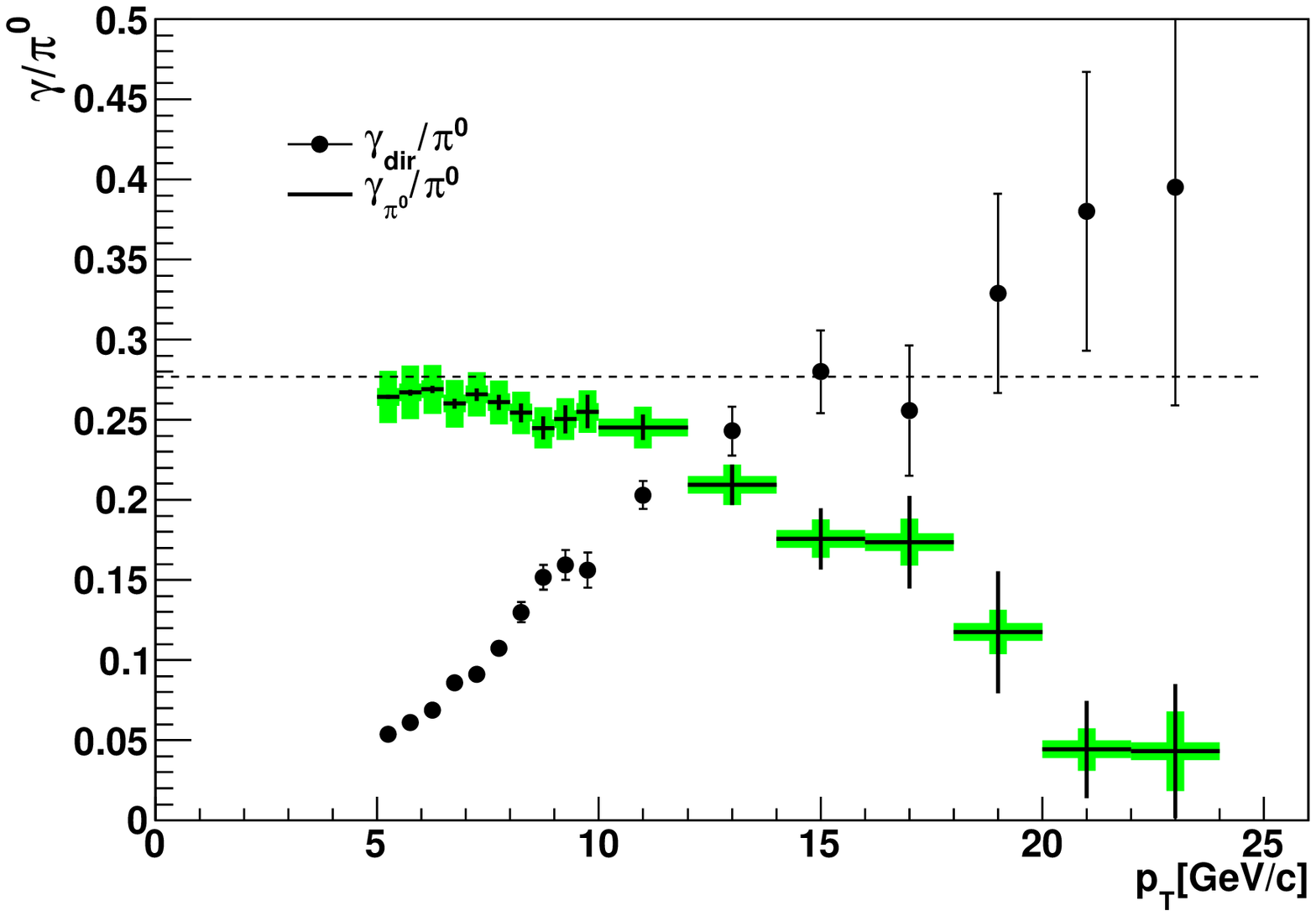}
\caption{(color online) 
  The filled circles ($\gamma_{\rm dir}/\pi^0$): the ratio of the direct photon cross 
  section to the $\pi^0$ cross section fit.
  The crosses ($\gamma_{\pi^{0}}/\pi^0$): the ratio of the background photon from 
  $\pi^0$ decay to the $\pi^0$ cross section fit.
  The $\pi^0$ cross section used for the denominator is the fit curve explained in 
  the text and shown in Fig. \ref{fig:crfit_pow1777_822}. The dotted line shows the 
  analytical expectation of $\gamma_{\pi^0}/\pi^0$. See the text for details.
  The systematic uncertainty on the $\pi^0$ photon contribution includes acceptance, 
  smearing, $\pi^0$ photon counting, and untagged $\pi^0$ photon probability. 
  (These uncertainties enter as A2, C, and D in Table \ref{table:syserror}).}
\label{fig:cr_gpi02pi0}
\end{figure}

To demonstrate the purity of the signal as a function of photon $p_T$,
ratios to $\pi^0$ spectrum are taken.  The measured $\pi^0$ cross
section \cite{Adare:2007dg} is parametrized by the form
$E\frac{d^3\sigma}{dp^3}$ [pb] $= 1.777\times 10^{10}\,p_T^{-8.22}$
as show in Fig. \ref{fig:crfit_pow1777_822}. The systematic
uncertainty shown with a band does not include the overall normalization
uncertainty.  Figure \ref{fig:cr_gpi02pi0} shows the ratio of both the
direct photon signal and photons from $\pi^0$ to the fit.  The dotted
line in the figure is at $2/(8.22-1)= 0.277$, which is the analytic
expectation for the ratio of $\pi^0$ decay photons to $\pi^0$ in case of a pure
power law behavior of the $\pi^0$.  The fraction of the direct photon
contribution gets higher as the transverse momentum increases. 
The contribution of photons from $\pi^0$s deviates from the analytic
expectation at higher $p_T$ because most of the merged clusters are
rejected from the photon sample. 
The data are systematically lower 
than the analytical line even at the lowest $p_T$. 
But as shown in Fig.\ref{fig:crfit_pow1777_822}, 
the systematic uncertainty of the fit constant is on the 
order of 10\%.  If this uncertainty is taken into account, 
the agreement is reasonable.


\begin{figure}[bh]
\includegraphics[width=1.0\linewidth]{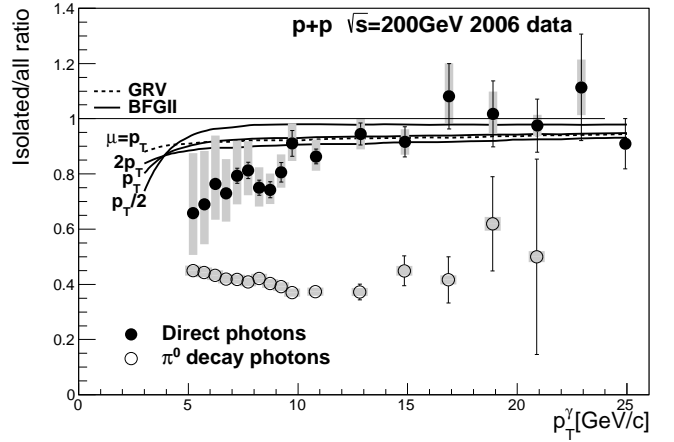}
\caption{The effect of isolation cut on direct photons and decay photons.
  Solid circles: Ratio of isolated direct photons to inclusive-direct
  photons. The statistical uncertainties are shown as black error bars
  and the systematic uncertainties are plotted as shaded bars. The
  solid and dashed curves are NLO pQCD calculations with three theory
  scales for BFGII \cite{Bourhis:1997yu} and one scale for the GRV
  \cite{Gluck:1992zx} parton to photon fragmentation functions. Open
  circles: Ratio of isolated photons from $\pi^0$ decays to all
  photons from $\pi^0$ decays.}
\label{fig:crisosub2}
\end{figure}

Figure \ref{fig:crisosub2} shows the ratio of isolated-direct photons to
inclusive-direct photons, and isolated over inclusive photons 
from $\pi^0$s. Since the background subtractions are done 
independently, this ratio
can exceed unity due to the uncertainty.  The data are
compared to two theoretical calculations, which include the same
isolation criteria.
The cone size of the isolation cut is larger than the area within the 
PHENIX central arm acceptance around the isolated photon candidate and 
leads to an underestimate of the energy in the cone.  This effect was not
corrected for; instead the same acceptance was included in the theory
calculation. However the theory calculation assumes no dead areas in the 
acceptance. 
The theory calculation varies at most by 2\% (90\% $\rightarrow$ 92\%), 
when the effect of the dead area is included.

The isolation cut causes a large suppression of photons from $\pi^0$.
This is expected as the $\pi^0$ is
accompanied with other fragmentation products.  At high $p_T$
($p_T>\sim 10$ GeV/$c$), the ratio of isolated direct photon to inclusive-direct photons
is typically more than 90\% and
matches the expectation from NLO pQCD calculations.  At low $p_T$, the
data are below the theory calculation, although generally they agree within 
the systematic uncertainty. Explanations of possible discrepancy were 
thought to be underlying event activity as
well as the contribution of photons from quark fragmentation, which
are not considered in the theory calculation.  However a study with an
event generator ({\sc pythia} tune A \cite{Sjostrand:1994ek}) did not show
any drop in the low $p_T$ region for the direct photons, while the
level of isolated photons from $\pi^0$ decays was well reproduced
\footnote{ In the previous paper \cite{Adler:2006yt}, it was claimed
  that the effect of underlying events was large. At that time, the
  Monte-Carlo calculation was done only for direct photon process and
  the ratio was scaled up to the NLO calculation.  For the present
  work, a full mixture of processes was generated, so the result can be
  directly compared with data, assuming that {\sc pythia} reproduces the
  physics correctly.}.

\section{Discussion}
\label{sec:discussion}
Figure~\ref{fig:spectrasub} shows good agreement between the data and
NLO pQCD calculations.  While the calculations at low $p_T$ seem low,
the correlated systematic uncertainties in the data are such that the
difference is not significant.

\begin{figure*}[t]
\includegraphics[width=0.75\linewidth]{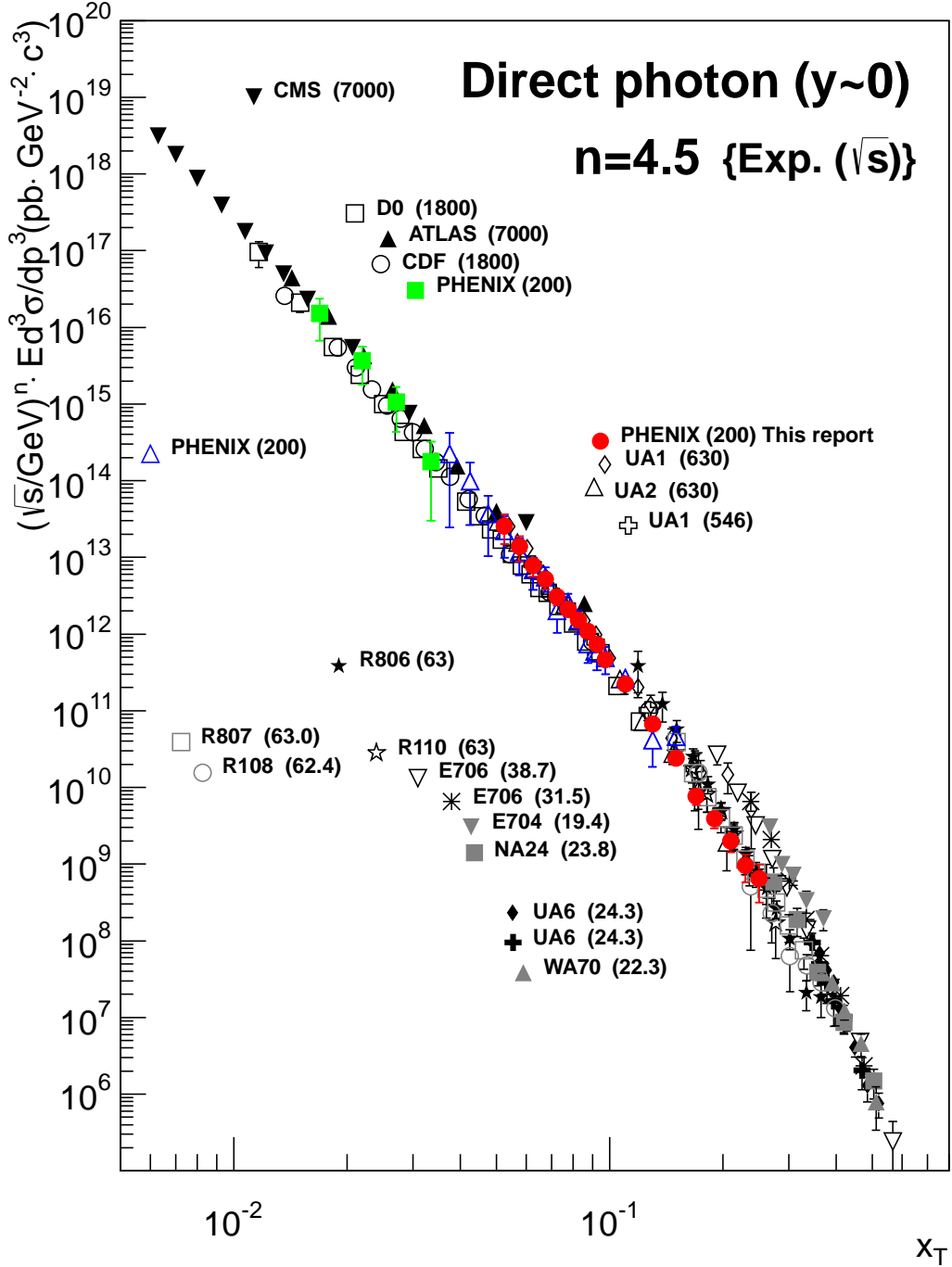}
\caption{(color online) Various direct photon cross section measurements in $p$$+$$p$ and
  $p+\bar{p}$ collisions scaled by $\sqrt{s}^{4.5}$ vs $x_T\equiv
  2p_T/\sqrt{s}$.  The legend shows the experiment
  and the center of mass energy [GeV] in parenthesis.
  Table~\protect\ref{table:xtdata} shows the references.
}
\label{fig:xtscaling}
\end{figure*}

\begin{table*}[ht]
\caption{Experimental data for Fig.~\protect\ref{fig:xtscaling}.  Points
  consistent with 0 are excluded from the plot. 
  Reference \cite{Vogelsang:1997cq} is a good review.}
\begin{ruledtabular}\begin{tabular}{llllllll}
System & Experiment & $\sqrt{s}$ & $E_T$ range & $\eta$ or $x_F$ range & iso cut & Data \\
       & [Ref.]     & [GeV]      &  [GeV]  &  & & points \\
\hline
$p$$+$$p$   & CMS  \cite{Khachatryan:2010fm} 	& 7000       & $22-210$ & $|\eta|<1.45$ & yes & 11 \\
$p$$+$$p$   & ATLAS \cite{Aad:2011tw}           & 7000       & $50-300$ & $|\eta|<0.6$ & yes & 8 \\
$p\bar{p}$ & D0 \cite{Abachi:1996qz}      & 1800       & $10.5-108.4$ & $|\eta|<0.9$ & yes & 23 \\
$p\bar{p}$ & CDF \cite{Abe:1994rra, Abe:1994rra_err}       & 1800       & $12.3-114.7$ & $|\eta|<0.9$ & yes & 16 \\
$p$$+$$p$   & PHENIX \cite{Adare:2009qk}        &  200       & $1.7-3.3$ & $|\eta|<0.35$ & no & 4 \\
$p$$+$$p$   & PHENIX (This report)             &  200       & $5.3-25$ & $|\eta|<0.25$& no & 18 \\
$p$$+$$p$   & PHENIX \cite{Adler:2006yt}       & 200        & $3.75-15$ & $|\eta|<0.25$ & no & 16 \\
$p\bar{p}$ & UA1 \cite{Albajar:1988im}   &  630       & $17-90$ & $|\eta|<0.8$ & yes & 16 \\
$p\bar{p}$ & UA1 \cite{Albajar:1988im}    &  546       & $17-46$ & $|\eta|<0.8$ & yes & 6 \\
$p\bar{p}$ & UA2 \cite{Alitti:1992hn}     &  630       & $15.9-82.3$ & $|\eta|<0.76$ & yes & 13 \\
$p$$+$$p$   & R110  \cite{Angelis:1989zv}      &  63       & $4.7-8.7$ & $|\eta|<0.8$ & yes & 7 \\
$p$$+$$p$   & R806 \cite{Anassontzis:1982gm}   & 63 & $3.75-11.50$ & $|\eta|<0.2$ & yes & 14 \\
$p$$+$$p$   & R807 \cite{Akesson:1989hp}       & 63 & $4.75-10.36$ & $|\eta|<0.7$ & yes & 11 \\
$p$$+$$p$   & R108 \cite{Angelis:1980yc}        & 62.4 & $5.37-12.44$ & $|\eta|<0.45$ & yes & 8 \\
$p$$+$$p$   & E706 \cite{Apanasevich:2004dr}   &  38.8       & $3.8-11$ & $-1. < \eta< 0.5$ & no & 9 \\
$p$$+$$p$   & E706 \cite{Apanasevich:2004dr}   &  31.6       & $3.8-9$ & $-0.75 < \eta< 0.75$ & no & 8 \\
$p$$+$$p$   & E704 \cite{Adams:1995gg}         &  19.4       & $2.6-3.6$ & $|x_F|<0.15$ & yes & 5 \\
$p$$+$$p$   & NA24 \cite{DeMarzo:1986vi}       &  23.8       & $3.3-6$ & $-0.65< \eta < 0.52$ & no & 5 \\
$p$$+$$p$   & WA70 \cite{Bonesini:1987bv}      &  23.0       & $4.1-5.7$ & $|x_F|<0.05$ & no & 5 \\
$p$$+$$p$   & UA6 \cite{Ballocchi:1998au}      &  24.3       & $4.2-6.3$ & $-0.1<\eta<0.9$ & no & 9 \\
$p\bar{p}$   & UA6 \cite{Sozzi:1993sm}  	&  24.3       & $4.2-5.7$ &$-0.1<\eta<0.9$ & no & 6 \\
\end{tabular}\end{ruledtabular}
\label{table:xtdata}
\end{table*}

Figure~\ref{fig:xtscaling} compiles this data and other measurements
of direct photons in $p$$+$$p$ or $p+\bar{p}$ collisions 
from both collider and fixed target experiments, over a broad range of
collision energy. 
Note that most of the collider data except for that of PHENIX apply
an isolation cut in their photon selection as listed in Table 
\ref{table:xtdata}. Also note that the PHENIX 
measurement \cite{Adare:2009qk} having the lowest $p_T$ points
uses the virtual-photon method. 
The cross sections are shown as a function of of
$x_T=2p_T/\sqrt{s}$ and scaled by the empirical value of
$\sqrt{s}^{n_{\rm eff}}$ with $n_{\rm eff}=4.5$ (Eq. \ref{eq:neffintro}).  
The effective power, $n_{\rm eff}$, is primarily
sensitive to the quantum exchange governing the reaction but also has
sensitivity to scale-breaking. For measurements of single particle or
single jet inclusive $p_T$ distributions, this $x_T$
scaling~\cite{Blankenbecler:1972cd,Cahalan:1974tp} provides a 
data driven test of whether pQCD or some other underlying subprocess
is at work, as well as providing a compact quantitative way to
describe the data using the effective index, $n_{\rm
  eff}(x_T,\sqrt{s})$
\begin{eqnarray}
 \nonumber  E \frac{d^3\sigma}{dp^3} & = & \frac{d^3\sigma}{p_T dp_T dy d\phi} \\
 \nonumber                           & = & {1 \over {p_T^{{n_{\rm eff}(x_T,\sqrt{s})}}}} F({p_T \over \sqrt{s}}) \\
                                    & = & {1\over {\sqrt{s}^{{\,n_{\rm eff}(x_T,\sqrt{s})}} } }G(x_T) \qquad,  
 \label{eq:neffintro}
\end{eqnarray}
where $E d^3\sigma/dp^3$ is the invariant cross section for inclusive
particle production with transverse momentum $p_T$ at c.m. energy
$\sqrt{s}$ and $x_T=2 p_T/\sqrt{s}$. It is important to emphasize that
the effective power, $n_{\rm eff}(x_T,\sqrt{s})$, is different from
the power $n$ of the invariant cross section at any given value of
$\sqrt{s}$.

For pure vector gluon exchange, or for QCD without evolution of
$\alpha_s$ and the structure and fragmentation functions, $n_{\rm
  eff}=4$ as in Rutherford scattering. However, due to the scale
breaking in QCD, the measured value of $n_{\rm eff}$ depends on the
$x_T$ value and the range of $\sqrt{s}$ used in the
computation~\cite{Cahalan:1974tp}.

For inclusive-direct photon production in $p$$+$$p$ collisions at
midrapidity, if we assume $x_1=x_2=x_T$
and $\langle\cos\theta^*\rangle = 0$, then from
Eq.\ref{eq:QCDComptonfull}, the $x_T$ scaled inclusive cross section
in pQCD is approximately:
\begin{eqnarray}
&& \nonumber \sqrt{s}^{n_{\rm eff}}\,E\frac{d^3\sigma}{dp^3} \\ &&
  \nonumber \propto \sqrt{s}^{(n_{\rm eff}-4)}\,\frac{x_T
    g_p(x_T,Q^2)\, F_{2p}(x_T,Q^2) \alpha_s(Q^2)}{x_T^4} \\ &&
  \propto \frac{x_T g_p(x_T)\, F_{2p}(x_T)}{x_T^4}
\end{eqnarray} 
where $x_T g_p$ is the gluon momentum distribution function in the
proton and $F_{2p}$ is the proton structure function measured in
DIS and we assume that the empirical value $n_{\rm eff}-4=0.5$ takes
account of the scale breaking effects.  The $x_T$ scaling of all the
available data, with some exception at low $\sqrt{s}$, is impressive.
As one goes to higher $x_T$, the power of the invariant cross section 
becomes softer. 
Figure \ref{fig:xtscaling} gives the same information as the agreement with the pQCD
calculations~\cite{Aurenche:2006vj} but in addition shows the validity
of pQCD directly from the data by a simple but powerful scaling
rule. 

\section{Summary and Conclusions}
\label{sec:summary}
The invariant differential cross section for the production of direct
photons in $p$$+$$p$ collisions at $\sqrt{s}=200$ GeV at midrapidity was
measured. It extends the $p_T$ reach up to 25 GeV/$c$. An NLO pQCD
calculation agrees well with the measurement, supporting the validity
of such calculations. 

The effect of an isolation cut on the direct-photon cross section 
was measured to be negligible ($<$10\%) in agreement with NLO 
theoretical calculations.  The isolation cut enhances the 
$g+q\rightarrow \gamma +q$ contribution and suppresses a possible 
background of single photons from bremsstrahlung or jet 
fragmentation. The main utility of the isolation cut is that it 
reduces the background of photons from hadronic decays by a 
significant factor of $\sim\!60$\%. Furthermore, the data are an 
important reference for interpreting direct-photon spectra in 
heavy-ion collisions.


\begin{acknowledgments}  

We thank the staff of the Collider-Accelerator and Physics
Departments at Brookhaven National Laboratory and the staff of
the other PHENIX participating institutions for their vital
contributions.  
We also thank Werner Vogelsang for providing calculations 
and for valuable, in-depth discussions.
We acknowledge support from the Office of Nuclear Physics in the
Office of Science of the Department of Energy,
the National Science Foundation, 
a sponsored research grant from Renaissance Technologies LLC, 
Abilene Christian University Research Council, 
Research Foundation of SUNY, 
and Dean of the College of Arts and Sciences, Vanderbilt University 
(U.S.A),
Ministry of Education, Culture, Sports, Science, and Technology
and the Japan Society for the Promotion of Science (Japan),
Conselho Nacional de Desenvolvimento Cient\'{\i}fico e
Tecnol{\'o}gico and Funda\c c{\~a}o de Amparo {\`a} Pesquisa do
Estado de S{\~a}o Paulo (Brazil),
Natural Science Foundation of China (P.~R.~China),
Ministry of Education, Youth and Sports (Czech Republic),
Centre National de la Recherche Scientifique, Commissariat
{\`a} l'{\'E}nergie Atomique, and Institut National de Physique
Nucl{\'e}aire et de Physique des Particules (France),
Ministry of Industry, Science and Tekhnologies,
Bundesministerium f\"ur Bildung und Forschung, Deutscher
Akademischer Austausch Dienst, and Alexander von Humboldt Stiftung (Germany),
Hungarian National Science Fund, OTKA (Hungary), 
Department of Atomic Energy and Department of Science and Technology (India),
Israel Science Foundation (Israel), 
National Research Foundation and WCU program of the 
Ministry Education Science and Technology (Korea),
Ministry of Education and Science, Russian Academy of Sciences,
Federal Agency of Atomic Energy (Russia),
VR and the Wallenberg Foundation (Sweden), 
the U.S. Civilian Research and Development Foundation for the
Independent States of the Former Soviet Union, 
the US-Hungarian Fulbright Foundation for Educational Exchange,
and the US-Israel Binational Science Foundation.

\end{acknowledgments}  

\appendix*
\section{}
Tables of the measured invariant differential cross section, 
the ratio of isolated to inclusive-direct photon, and systematic uncertainties.

\clearpage

\begin{table}[ht]
\caption{Cross section of midrapidity inclusive-direct photon production in $p$$+$$p$ 
collisions at $\sqrt{s}=200$ GeV as a function of transverse momentum ($p_T$).
Asymmetric statistical uncertainties occur in $p_{T}$ bins with no tagged
$\pi^{0}$ counts.}
\begin{ruledtabular}\begin{tabular}{lllll}
$p_T$ & $Ed^3\sigma/dp^3$ & Stat- & Stat+ & Syst \\
$[$GeV/$c]$ & $[pb\cdot GeV^{-2}\cdot c^3]$ & & & \\
\hline
 5.25 & 1.14e+03 & 3.04e+01 & 3.04e+01 & 4.78e+02 \\
 5.75 & 6.13e+02 & 1.92e+01 & 1.92e+01 & 2.21e+02 \\
 6.25 & 3.48e+02 & 1.27e+01 & 1.27e+01 & 1.01e+02 \\
 6.75 & 2.31e+02 & 8.50e+00 & 8.50e+00 & 6.24e+01 \\
 7.25 & 1.36e+02 & 6.12e+00 & 6.12e+00 & 3.13e+01 \\
 7.75 & 9.29e+01 & 4.41e+00 & 4.41e+00 & 1.95e+01 \\
 8.25 & 6.70e+01 & 3.22e+00 & 3.22e+00 & 1.34e+01 \\
 8.75 & 4.83e+01 & 2.45e+00 & 2.45e+00 & 9.18e+00 \\
 9.25 & 3.21e+01 & 1.89e+00 & 1.89e+00 & 6.10e+00 \\
 9.75 & 2.04e+01 & 1.46e+00 & 1.46e+00 & 3.68e+00 \\
11.00 & 9.81e+00 & 4.23e-01 & 4.23e-01 & 1.67e+00 \\
13.00 & 2.97e+00 & 1.89e-01 & 1.89e-01 & 4.75e-01 \\
15.00 & 1.06e+00 & 9.85e-02 & 9.85e-02 & 1.69e-01 \\
17.00 & 3.38e-01 & 5.51e-02 & 5.51e-02 & 5.42e-02 \\
19.00 & 1.73e-01 & 3.37e-02 & 3.37e-02 & 2.77e-02 \\
21.00 & 8.82e-02 & 2.06e-02 & 2.01e-02 & 1.50e-02 \\
23.00 & 4.22e-02 & 1.52e-02 & 1.41e-02 & 7.18e-03 \\
25.00 & 2.87e-02 & 1.41e-02 & 9.07e-03 & 4.30e-03 \\
\end{tabular}\end{ruledtabular}
\label{table:hg_sub}

\caption{Ratio of isolated/inclusive-direct photon (Fig. \ref{fig:crisosub2}).
Upper(+) and lower bounds(-) on systematics can be different, and are listed separately.}
\begin{ruledtabular}\begin{tabular}{lllll}
$<p_T>$ & Ratio & Stat & Syst+ & Syst- \\
$[$GeV/$c]$ & & & & \\
\hline
 5.23 & 0.658 & 0.014 & 0.217 & 0.151\\
 5.73 & 0.690 & 0.017 & 0.193 & 0.145\\
 6.23 & 0.764 & 0.022 & 0.176 & 0.130\\
 6.73 & 0.730 & 0.021 & 0.124 & 0.102\\
 7.23 & 0.793 & 0.027 & 0.127 & 0.103\\
 7.73 & 0.813 & 0.029 & 0.106 & 0.089\\
 8.23 & 0.750 & 0.028 & 0.074 & 0.067\\
 8.74 & 0.742 & 0.029 & 0.059 & 0.052\\
 9.24 & 0.806 & 0.035 & 0.064 & 0.056\\
 9.74 & 0.911 & 0.047 & 0.073 & 0.064\\
10.83 & 0.863 & 0.027 & 0.060 & 0.052\\
12.85 & 0.945 & 0.040 & 0.057 & 0.057\\
14.87 & 0.916 & 0.055 & 0.046 & 0.046\\
16.89 & 1.082 & 0.118 & 0.119 & 0.097\\
18.90 & 1.017 & 0.119 & 0.081 & 0.071\\
20.91 & 0.975 & 0.096 & 0.039 & 0.039\\
22.92 & 1.114 & 0.193 & 0.100 & 0.100\\
24.92 & 0.909 & 0.091 & 0.000 & 0.000\\
\end{tabular}\end{ruledtabular}
\label{table:ratio_gamma}
\end{table}

\begin{table}[ht]
\caption{\label{table:ratio_pi0}
Ratio of isolated/inclusive photon from $\pi^0$ (Fig. \ref{fig:crisosub2}).}
\begin{ruledtabular}\begin{tabular}{llll}
$<p_T>$ & ratio & Stat & Syst+- \\
$[$GeV/$c]$ & & & \\
\hline
 5.22 & 0.450 & 0.003 & 0.018 \\
 5.72 & 0.443 & 0.003 & 0.018 \\
 6.23 & 0.433 & 0.004 & 0.017 \\
 6.73 & 0.419 & 0.006 & 0.017 \\
 7.23 & 0.418 & 0.007 & 0.017 \\
 7.73 & 0.409 & 0.009 & 0.016 \\
 8.23 & 0.422 & 0.012 & 0.017 \\
 8.73 & 0.403 & 0.014 & 0.016 \\
 9.23 & 0.392 & 0.017 & 0.016 \\
 9.73 & 0.370 & 0.020 & 0.015 \\
10.79 & 0.373 & 0.015 & 0.015 \\
12.82 & 0.372 & 0.029 & 0.015 \\
14.84 & 0.449 & 0.054 & 0.018 \\
16.86 & 0.416 & 0.083 & 0.017 \\
18.88 & 0.619 & 0.171 & 0.025 \\
20.89 & 0.500 & 0.354 & 0.020 \\
\end{tabular}\end{ruledtabular}


\caption{\label{table:syserrorisoratio}
Components of systematic uncertainties on isolated/inclusive-direct photon ratio 
(expressed as a percentage of the center value).
Note that the upper and lower variations are treated separately.}
\begin{ruledtabular}\begin{tabular}{rrrrr}
$\langle p_T \rangle$ & pi0\_tagging & pi0\_miss\_ratio & non\_vertex & other\_hadron \\ 
\hline
  5.2 &  25.1 / -16.6 &  20.0 / -14.3 &   4.0 /  -3.5 &   7.7 /  -5.9\\
  5.7 &  21.6 / -15.0 &  16.4 / -12.4 &   3.8 /  -3.3 &   7.1 /  -5.7\\
  6.2 &  18.7 / -13.5 &  10.8 /  -8.8 &   3.6 /  -3.2 &   6.9 /  -5.7\\
  6.7 &  13.8 / -10.8 &   7.6 /  -6.6 &   2.8 /  -2.5 &   5.2 /  -4.5\\
  7.2 &  13.1 / -10.3 &   5.7 /  -5.1 &   2.8 /  -2.5 &   5.3 /  -4.5\\
  7.7 &  10.7 /  -8.8 &   4.5 /  -4.1 &   2.4 /  -2.2 &   4.5 /  -4.0\\
  8.2 &   8.4 /  -7.2 &   3.4 /  -3.2 &   1.8 /  -1.7 &   3.4 /  -3.1\\
  8.7 &   6.8 /  -6.0 &   2.7 /  -2.6 &   1.5 /  -1.4 &   2.9 /  -2.6\\
  9.2 &   6.6 /  -5.8 &   2.6 /  -2.5 &   1.5 /  -1.4 &   3.0 /  -2.7\\
  9.7 &   7.0 /  -6.1 &   2.6 /  -2.5 &   1.7 /  -1.6 &   3.4 /  -3.1\\
 10.8 &   5.3 /  -4.8 &   2.1 /  -2.0 &   1.3 /  -1.2 &   4.0 /  -3.6\\
 12.9 &   3.8 /  -3.6 &   1.9 /  -1.8 &   1.0 /  -0.9 &   4.6 /  -4.1\\
 14.9 &   2.9 /  -2.7 &   1.8 /  -1.7 &   0.7 /  -0.7 &   4.0 /  -3.6\\
 16.9 &   3.5 /  -3.3 &   3.0 /  -2.9 &   0.9 /  -0.9 &  10.4 /  -8.2\\
 18.9 &   2.0 /  -1.9 &   2.6 /  -2.5 &   0.4 /  -0.4 &   7.3 /  -5.9\\
 20.9 &   0.9 /  -0.8 &   2.8 /  -2.7 &   0.2 /  -0.2 &   3.1 /  -2.8\\
 22.9 &   1.0 /  -1.0 &   7.2 /  -6.3 &   0.3 /  -0.3 &   6.5 /  -5.7\\
 24.9 &   0.0 /   0.0 &   0.0 /   0.0 &  -0.0 /  -0.0 &   0.0 /   0.0\\
\end{tabular}\end{ruledtabular}

\end{table}

\begingroup
\squeezetable
\begin{table*}[ht]
\caption{\label{table:syserror}
Systematic uncertainties of inclusive-direct photon measurement. 
The percentage uncertainties of each component is shown: 
(A) the global factor (quadrature sum of components A1, A2, and A3), 
(B) inclusive photon counts, 
(C) $\pi^0$ tagging, (D) the factor for the total $\pi^0$ counts, and (E) the factor for the hadron contribution 
other than $\pi^0$) and 
the contribution to the direct photon signal (A*1, B*W, C*(W-1), D*(W-1), E*(W-1)). 
W is the ratio of $N_{\rm incl}/N_{\rm dir}$.
Individual components are 
(A1) Energy scale error transformed to the cross section, 
(A2) Acceptance and smearing, 
(A3) $\sigma_{\rm BBC}$ and BBC trigger bias,
(B1) nonvertex, neutral hadron subtraction, 
(C1) $\pi^0$ combinatorial background subtraction, 
(C2) loss for conversion and Dalitz decay, 
(D1) $E_{\rm min}$ calibration, 
(D2) input $\pi^0$ spectrum in the MC, 
(D3) $\pi^0$ merge model (correction for the complete merging), 
(D4) $\pi^0$ merge model (cluster shape parametrization in the MC), 
(D5) Geometry and trigger mask,
(E1) Ratio of all hadronic decay to $\pi^0$ contribution,
(E2) Isolation with own decay partner, and 
(E3) $\pi^0$ merge correction for other hadron contribution.}
\begin{ruledtabular}\begin{tabular}{ r r r r r r r r r r r r r r r r r r r r  r }
$p_T$ & 1/W & A1 & A2 & A3 & A*1 & B1 & B*W & C1 & C2 & C*(W-1) & D1 & D2 & D3 & D4 & D5 & D*(W-1) & E1 & E2 & E*(W-1) & total \%  \\ 
\hline
5.25 & 0.13 & 10 & 3 & 10 & 14.46 & 1 & 7.49 & 3 & 1 & 20.53    & 4 & 2 & 0 & 0 & 1 & 29.74 & 2 & 0 & 12.98 & 41.71  \\ 
5.75 & 0.15 & 10 & 3 & 10 & 14.46 & 1 & 6.47 & 3 & 1 & 17.31    & 4 & 2 & 0 & 0 & 1 & 25.08 & 2 & 0 & 10.94 & 36.04  \\ 
6.25 & 0.18 & 10 & 3 & 10 & 14.46 & 1 & 5.7 & 3 & 1 & 14.86     & 3 & 2 & 0 & 0 & 1 & 17.58 & 2 & 0 & 9.4 & 29.32  \\ 
6.75 & 0.2 & 10 & 3 & 10 & 14.46 & 1 & 5.09 & 3 & 1 & 12.93    & 3 & 2 & 0 & 0 & 1 & 15.3 & 2 & 0 & 8.18 & 26.52  \\ 
7.25 & 0.22 & 10 & 3 & 10 & 14.46 & 1 & 4.6 & 3 & 1 & 11.38     & 2 & 2 & 0 & 0 & 1 & 10.79 & 2 & 0 & 7.2 & 22.97  \\ 
7.75 & 0.24 & 10 & 3 & 10 & 14.46 & 1 & 4.19 & 3 & 1 & 10.1     & 2 & 2 & 0 & 0 & 1 & 9.58 & 2 & 0 & 6.39 & 21.47  \\ 
8.25 & 0.26 & 10 & 3 & 10 & 14.46 & 1 & 3.85 & 3 & 1 & 9.02     & 2 & 2 & 0 & 0 & 1 & 8.56 & 2 & 0 & 5.71 & 20.28  \\ 
8.75 & 0.28 & 10 & 3 & 10 & 14.46 & 1 & 3.57 & 3 & 1 & 8.11     & 2 & 2 & 0 & 0 & 1 & 7.7 & 2 & 0 & 5.13 & 19.31  \\ 
9.25 & 0.3 & 10 & 3 & 10 & 14.46 & 1 & 3.32 & 3 & 1 & 7.33     & 2 & 2 & 0 & 0 & 1 & 6.95 & 2 & 0 & 4.63 & 18.53  \\ 
9.75 & 0.32 & 10 & 3 & 10 & 14.46 & 1 & 3.1 & 3 & 1 & 6.64      & 2 & 2 & 0 & 0 & 1 & 6.3 & 2 & 1 & 4.7 & 18.01  \\ 
11 & 0.38 & 10 & 3 & 10 & 14.46 & 1 & 2.67 & 3 & 1 & 5.27       & 2 & 2 & 0 & 1 & 1 & 5.27 & 2 & 2 & 4.71 & 17.14  \\ 
13 & 0.46 & 10 & 3 & 10 & 14.46 & 1 & 2.18 & 3 & 1 & 3.73       & 2 & 2 & 0 & 3 & 1 & 5 & 2 & 3 & 4.25 & 16.45  \\ 
15 & 0.54 & 10 & 3 & 10 & 14.46 & 1 & 1.84 & 3 & 1 & 2.66       & 2 & 2 & 1.7 & 4 & 1 & 4.44 & 2 & 4 & 3.76 & 15.92  \\ 
17 & 0.63 & 10 & 3 & 10 & 14.46 & 1 & 1.59 & 3 & 1 & 1.88       & 2 & 2 & 3.1 & 6 & 1 & 4.4 & 2 & 6 & 3.76 & 15.77  \\ 
19 & 0.71 & 10 & 3 & 10 & 14.46 & 1 & 1.41 & 3 & 1 & 1.29       & 2 & 2 & 6.7 & 8 & 1 & 4.41 & 2 & 8 & 3.35 & 15.6  \\ 
21 & 0.8 & 10 & 3 & 10 & 14.46 & 1 & 1.26 & 3 & 1 & 0.82       & 2 & 2 & 27.4 & 10 & 1 & 7.56 & 2 & 10 & 2.63 & 16.59  \\ 
23 & 0.88 & 10 & 3 & 10 & 14.46 & 1 & 1.14 & 3 & 1 & 0.44       & 2 & 2 & 56.3 & 12 & 1 & 7.93 & 2 & 12 & 1.67 & 16.62  \\ 
25 & 0.96 & 10 & 3 & 10 & 14.46 & 1 & 1.04 & 3 & 1 & 0.12       & 2 & 2 & 102.8 & 14 & 1 & 3.99 & 2 & 13 & 0.51 & 15.04  \\ 
\end{tabular}\end{ruledtabular}

%


\caption{Systematic uncertainties of isolated-direct photon measurement. 
The percentage uncertainties of each component are as given in the Table~\protect\ref{table:syserror} caption,
except that the contributions to the direct photon signal are A*1, B*W0, C*W3, C*W2, D*W1, and E*W1. 
W0, W1, W2, and W3 are defined in Eq.\ref{eqn:isolation_sys}.
Individual components A1 through E3 are also as given in the Table~\protect\ref{table:syserror} caption.}
\label{table:isosyserror}
\begin{ruledtabular}\begin{tabular}{ r r r r r r r r r r r r r r r r r r r r r r r r r r }
$p_T$ & W0 & W1 & W2 & W3  & A1 & A2 & A3 & A*1 & B1 & B*W0 & C1 & C2 & C*W3 & C*W2 & D1 & D2 & D3 & D4 & D5 & D*W1 & E1 & E2 & E3 & E*W1 & Total \% \\ 
\hline
5.25 & 3.69 & 4.16 &  0.1  & 2.62 & 10 & 3 & 10 & 14.46 & 1 & 3.69 & 3 & 1 & 8.3  & 0.3	 & 4 & 2 & 0	  & 0  & 1 & 19.07 & 2 & 0.1 & 0  & 8.33 & 26.92 \\
5.75 & 3.09 & 3.27 &  0.09 & 1.98 & 10 & 3 & 10 & 14.46 & 1 & 3.09 & 3 & 1 & 6.27 & 0.3	 & 4 & 2 & 0	  & 0  & 1 & 14.99 & 2 & 0.1 & 0  & 6.55 & 22.92 \\
6.25 & 2.65 & 2.62 &  0.09 & 1.53 & 10 & 3 & 10 & 14.46 & 1 & 2.65 & 3 & 1 & 4.83 & 0.29 & 3 & 2 & 0	  & 0  & 1 & 9.79  & 2 & 0.3 & 0  & 5.29 & 19.06 \\
6.75 & 2.32 & 2.13 &  0.09 & 1.2  & 10 & 3 & 10 & 14.46 & 1 & 2.32 & 3 & 1 & 3.8  & 0.29 & 3 & 2 & 0	  & 0  & 1 & 7.96  & 2 & 0.3 & 0  & 4.3  & 17.63 \\
7.25 & 2.08 & 1.76 &  0.09 & 0.96 & 10 & 3 & 10 & 14.46 & 1 & 2.08 & 3 & 1 & 3.04 & 0.28 & 2 & 2 & 0	  & 0  & 1 & 5.27  & 2 & 0.4 & 0  & 3.58 & 16.23 \\
7.75 & 1.89 & 1.47 &  0.09 & 0.78 & 10 & 3 & 10 & 14.46 & 1 & 1.89 & 3 & 1 & 2.48 & 0.28 & 2 & 2 & 0	  & 0  & 1 & 4.42  & 2 & 0.5 & 0  & 3.04 & 15.73 \\
8.25 & 1.75 & 1.25 &  0.09 & 0.65 & 10 & 3 & 10 & 14.46 & 1 & 1.75 & 3 & 1 & 2.06 & 0.27 & 2 & 2 & 0	  & 0  & 1 & 3.76  & 2 & 0.6 & 0  & 2.61 & 15.41 \\
8.75 & 1.64 & 1.08 &  0.08 & 0.55 & 10 & 3 & 10 & 14.46 & 1 & 1.64 & 3 & 1 & 1.73 & 0.27 & 2 & 2 & 0	  & 0  & 1 & 3.24  & 2 & 0.7 & 0  & 2.29 & 15.18 \\
9.25 & 1.56 & 0.94 &  0.08 & 0.47 & 10 & 3 & 10 & 14.46 & 1 & 1.56 & 3 & 1 & 1.48 & 0.26 & 2 & 2 & 0	  & 0  & 1 & 2.82  & 2 & 0.7 & 0  & 1.99 & 15.02 \\
9.75 & 1.49 & 0.83 &  0.08 & 0.41 & 10 & 3 & 10 & 14.46 & 1 & 1.49 & 3 & 1 & 1.28 & 0.26 & 2 & 2 & 0	  & 0  & 1 & 2.49  & 2 & 0.9 & 0  & 1.82 & 14.91 \\
11   & 1.37 & 0.63 &  0.08 & 0.3  & 10 & 3 & 10 & 14.46 & 1 & 1.37 & 3 & 1 & 0.95 & 0.24 & 2 & 2 & 0	  & 1  & 1 & 2	   & 2 & 1   & 1  & 1.55 & 14.77 \\
13   & 1.27 & 0.46 &  0.07 & 0.21 & 10 & 3 & 10 & 14.46 & 1 & 1.27 & 3 & 1 & 0.66 & 0.23 & 2 & 2 & 0	  & 3  & 1 & 1.94  & 2 & 1.5 & 2  & 1.46 & 14.73 \\
15   & 1.22 & 0.36 &  0.07 & 0.16 & 10 & 3 & 10 & 14.46 & 1 & 1.22 & 3 & 1 & 0.51 & 0.21 & 2 & 2 & 1.7	  & 4  & 1 & 1.89  & 2 & 2.3 & 3  & 1.53 & 14.72 \\
17   & 1.2  & 0.29 &  0.06 & 0.13 & 10 & 3 & 10 & 14.46 & 1 & 1.2  & 3 & 1 & 0.41 & 0.19 & 2 & 2 & 3.1	  & 6  & 1 & 2.12  & 2 & 3.2 & 4  & 1.58 & 14.75 \\
19   & 1.18 & 0.22 &  0.06 & 0.1  & 10 & 3 & 10 & 14.46 & 1 & 1.18 & 3 & 1 & 0.32 & 0.18 & 2 & 2 & 6.7	  & 8  & 1 & 2.43  & 2 & 4.3 & 6  & 1.71 & 14.81 \\
21   & 1.15 & 0.16 &  0.05 & 0.07 & 10 & 3 & 10 & 14.46 & 1 & 1.15 & 3 & 1 & 0.23 & 0.17 & 2 & 2 & 27.4	  & 10 & 1 & 4.73  & 2 & 5.4 & 8  & 1.59 & 15.34 \\
23   & 1.12 & 0.1  &  0.05 & 0.04 & 10 & 3 & 10 & 14.46 & 1 & 1.12 & 3 & 1 & 0.14 & 0.15 & 2 & 2 & 56.3	  & 12 & 1 & 5.85  & 2 & 6.5 & 9  & 1.14 & 15.68 \\
25   & 1.09 & 0.05 &  0.05 & 0.02 & 10 & 3 & 10 & 14.46 & 1 & 1.09 & 3 & 1 & 0.07 & 0.14 & 2 & 2 & 102.8  & 14 & 1 & 5.48  & 2 & 7.5 & 10 & 0.67 & 15.51 \\
\end{tabular}\end{ruledtabular}

\end{table*}
\endgroup


\clearpage



\end{document}